\documentclass[prd,aps,eqsecnum,floatfix,nofootinbib,preprint,tightenlines]{revtex4}

\usepackage{latexsym}
\usepackage{graphicx}
\usepackage{multirow}
\usepackage{color}
\usepackage{amsmath}
\usepackage{hyperref}  

\def\floatcaption#1#2{ \caption{#2 \label{#1}} }

\def\NON{\nonumber\\}
\def\bibi{\bibitem}

\def\a{\alpha}
\def\b{\beta}

\def\d{\delta}
\def\e{\epsilon}                
\def\f{\phi}                    

\def\h{\eta}

\def\l{\lambda}
\def\m{\mu}

\def\p{\pi}                     
\def\th{\theta}                  
\def\r{\rho}                    
\def\s{\sigma}                  

\def\x{\xi}
\def\z{\zeta}

\def\G{\Gamma}

\def\L{\Lambda}

\def\S{\Sigma}

\def\ca{{\cal A}}
\def\cb{{\cal B}}
\def\cc{{\cal C}}
\def\cd{{\cal D}}

\def\cl{{\cal L}}

\def\cp{{\cal P}}

\def\bo{\raisebox{-.4ex}{\large$\Box$}}                 
\def\cbo{{\,\raise-.15ex\Sc [\,}}                       

\def\svev#1{\left\langle #1\right\rangle}       

\def\ddt#1{{\buildrel {\hbox{\LARGE .\kern-2pt.}} \over {#1}}}

\def\ie{\mbox{\it i.e.}}
\def\eg{\mbox{\it e.g.}}

\def\tr{{\rm tr}\,}

\def\half{{1\over 2}}

\def\ttl#1{{\it #1}}
\def\ttl#1{}

\long \def \blockcomment #1\endcomment{}

\def\cdt{\!\cdot}

\def\bc{\overline{C}}

\def\tA{\tilde{A}}
\def\tG{\tilde{G}}
\def\tg{\tilde{g}}
\def\tL{\tilde{\L}}
\def\tl{\tilde{\l}}
\def\tX{\widetilde{X}}
\def\tT{\tilde{T}}

\def\vn{{\vec{n}}}
\def\vm{{\vec{m}}}
\def\vv{{\vec{v}}}
\def\hv{{\hat{v}}}

\def\vc{{\vec C}}
\def\vbc{{{\vec{\bc}}}}

\begin{document}

\begin{center}
\vspace{10mm}
{\large\bf
Dimensional transmutation in the longitudinal sector of\\[3mm]
equivariantly gauge-fixed Yang--Mills theory
}
\\[12mm]
Maarten Golterman$^a$
and Yigal Shamir$^b$
\\[8mm]
{\small\it
$^a$Department of Physics and Astronomy\\
San Francisco State University, San Francisco, CA 94132, USA}%
\\[5mm]
{\small\it $^b$Raymond and Beverly Sackler School of Physics and Astronomy\\
Tel-Aviv University, Ramat~Aviv,~69978~ISRAEL}%
\\[10mm]
 {ABSTRACT}
\\[2mm]
\end{center}

\begin{quotation}
We study the pure-gauge sector of an $SU(N)$ gauge theory, equivariantly
gauge fixed to $SU(N-1)\times U(1)$, which is an asymptotically free
non-linear sigma model in four dimensions.
We show that dimensional transmutation takes place in the large-$N$ limit,
and elaborate on the relevance of this result for a speculative scenario
in which the strong longitudinal dynamics gives rise to a novel
Higgs--Coulomb phase.
\end{quotation}

\newpage
\section{\label{Intro} Introduction}
While in continuum Yang--Mills theories gauge fixing is needed in order to define them through
weak-coupling perturbation theory, the standard lattice formulation of
these theories is well defined
without any gauge fixing, thanks to compactness of the group variables.
Nevertheless, also in lattice QCD gauge fixing is often used, for
practical reasons.   It then usually refers to the procedure of first
generating an ensemble of gauge field configurations
using a gauge-invariant action, and then applying to each configuration
a gauge transformation that rotates it to some gauge of choice.
This generalizes to any gauge theory without fermions, or with a vector-like
fermion content.

The non-perturbative definition of chiral gauge theories using the lattice
is not an entirely solved problem \cite{LChGT}.
Some years ago we proposed a construction where
the chiral gauge symmetry is explicitly broken on the lattice, but is
recovered in the continuum limit for any anomaly-free and asymptotically free
chiral gauge theory \cite{nachgt}.  An essential ingredient of our proposal
is the inclusion of a local gauge-fixing action
as part of the very definition of the lattice theory.  In contrast to the above-mentioned
gauge-fixing procedure often used in lattice QCD, the gauge-fixed theory is
defined by a single local action on the lattice.

Lattice gauge-fixing actions can be studied on their own right, and they have
been studied in the past.
Neuberger proved that insisting on BRST symmetry, as in
the usual continuum treatment, leads to an impasse.
The partition function itself, as well as all (unnormalized) expectation
values of gauge-invariant operators, vanish \cite{HNnogo}.
The problem can be avoided if, instead of the entire gauge group $G$,
only a coset $G/H$ is gauge fixed, leaving
the local $H$-invariance unfixed \cite{MS1,nachgt}, for appropriate
choices of the subgroup $H$.
In such ``equivariantly'' gauge-fixed theories
one expects an invariance theorem to apply: Finite-volume correlation functions
of gauge-invariant observables should be well-defined, and equal
to those of the original lattice gauge theory without any gauge fixing.
Using  equivariant BRST symmetry (eBRST, for short),
we proved the invariance theorem for $G=SU(N)$ gauge theories
in the case that $H$ is the Cartan
subgroup \cite{MS1}, as well as in the case that it is a maximal
subgroup \cite{GSMF}.%
\footnote{
  We believe that the proof can be generalized to any subgroup $H$
  containing the Cartan subgroup.
}
An essential ingredient of the proof is the exact invariance of the
theory under eBRST symmetry.

An equivariantly gauge-fixed theory depends on two coupling constants.
The dynamics of the transverse sector is controlled by the usual
gauge coupling $g$.  The dynamics of all other degrees of freedom,
to which we will refer as the longitudinal sector, is controlled by
another coupling $\tg=\sqrt\x g$, where $\x$ is the gauge-fixing parameter.

Like the transverse coupling $g$, also the longitudinal coupling $\tg$
is asymptotically free \cite{GSb}.  This raises many interesting questions.
First, can dimensional transmutation take place in the longitudinal sector,
as it does in the transverse sector?  The solution of the one-loop
renormalization-group equation answers this question
in the affirmative \cite{GSb}:
Choose initial conditions for the running couplings such that,
at some high-energy scale, the renormalized couplings $g_r$ and $\tg_r$
are both weak (so that
perturbation theory is applicable), but $\tg_r$ is much bigger than $g_r$.
Then there is a low-energy scale $\tL$ where
the running longitudinal coupling $\tg_r$
becomes strong while $g_r$ is still weak.
Stated differently, if $\L$ is the usual
confinement scale where the one-loop running coupling $g_r$ becomes strong,
then, for the initial conditions above, the longitudinal coupling
will become strong at a scale $\tL\gg\L$.

An intriguing question is whether a strong dynamics in the longitudinal
sector can possibly have any effect on the gauge-invariant sector
of the theory.  In Ref.~\cite{GSMF} we studied this question for an $SU(2)$
lattice gauge theory equivariantly gauge fixed to $U(1)$,
using a combination of strong-coupling and mean-field techniques.
While mean field is not a systematic approximation, it can provide
valuable clues as to what might happen.  Our conclusion was that
a part of the bare-parameter phase diagram where $\tg_0/g_0\gg 1$ may
belong to a novel phase that resembles the broken phase
of the Georgi--Glashow model \cite{GG,LS}.  In this phase, two of the gauge
bosons become massive.  The third gauge boson, a photon, remains massless.
In continuum field-theory parlance, the $SU(2)$ gauge symmetry is spontaneously
broken to $U(1)$.  Stated more rigorously, in the non-perturbative lattice theory, the existence
of a massless photon distinguishes the Higgs--Coulomb phase from the
confinement phase.

At face value, this is a paradoxical claim.  It appears to contradict
the invariance theorem, according to which gauge invariant observables
must be independent of the longitudinal (lattice) coupling $\tg_0$.
In fact, there is no contradiction.
Introducing a small breaking of eBRST symmetry into the theory
(which we will generically refer to as a ``seed''),  we then first take
the infinite-volume limit, and only after that proceed to turn off the seed.
As explained in detail in Ref.~\cite{GSMF},
with this order of limits, the invariance theorem no longer holds.
It then becomes a dynamical question whether or not the consequences
of the theorem will be recovered in the thermodynamical limit.

The non-perturbative physics of the tentative Higgs--Coulomb phase is driven
by the longitudinal sector, while the transverse sector remains weakly coupled.
We can isolate the longitudinal sector by freezing out the transverse
degrees of freedom.  The resulting theory is what we will refer to as the ``reduced model,'' and
corresponds to the $g_0=0$ boundary of phase diagram.
The configuration space of the reduced model
is the trivial orbit $U_\m(x)=\phi(x)\phi^\dagger(x+\m)$, with a group-valued
scalar field $\phi(x)\in G$ taking the place of the lattice
gauge field $U_\m(x)$.
Like the original coset gauge-fixed theory,
the reduced model has local $H$ invariance,
which implies that $\phi(x)$ actually lives in the coset $G/H$.

The reduced model ``inherits'' a global $G$ symmetry, and the
tentative novel phase is characterized by the spontaneous breaking
of the global $G$ symmetry to its $H$ subgroup.  When the transverse
gauge field will be turned back on, the Nambu--Goldstone bosons
(NGBs) of this
symmetry breaking will be eaten by the $G/H$ coset gauge fields,
which in turn will become massive.  Thus, if indeed the reduced model
has a $G\to H$ broken-symmetry phase, it is all but natural that it will
become the boundary of a Higgs--Coulomb phase of the full theory.

The eBRST invariance of the reduced model makes it a topological field
theory.%
\footnote{
  The proof of this result is closely related to the proof of
  the invariance theorem \cite{nachgt}.
}
The finite-volume partition function is independent
of the single coupling $\tg_0$.
The question arises whether such a topological field theory can
nevertheless generate
a non-trivial effective potential, without which no spontaneous
symmetry breaking can take place.  In Ref.~\cite{GSMF} we studied this question
in a zero-dimensional toy model, and concluded that the answer is yes.
A key feature common to the original eBRST gauge-fixed theory,
its reduced model,
and the toy model of Ref.~\cite{GSMF}, is that the measure can be both positive
and negative.  The inclusion of a seed thus serves two functions.
First, as usual, it tilts the effective potential,
choosing a preferred orientation for the broken-symmetry vacuum.
But, in addition, by breaking the eBRST symmetry
the seed violates the fragile balance between
configurations with a positive and with a negative measure,
which was protected by this symmetry, so that both the toy model
and the reduced model are no longer topological field theories.
The conclusion is that it is indeed possible for an effective potential
to exist, and to induce a non-trivial, broken symmetry vacuum.

\begin{figure}
\includegraphics*[width=7.0cm]{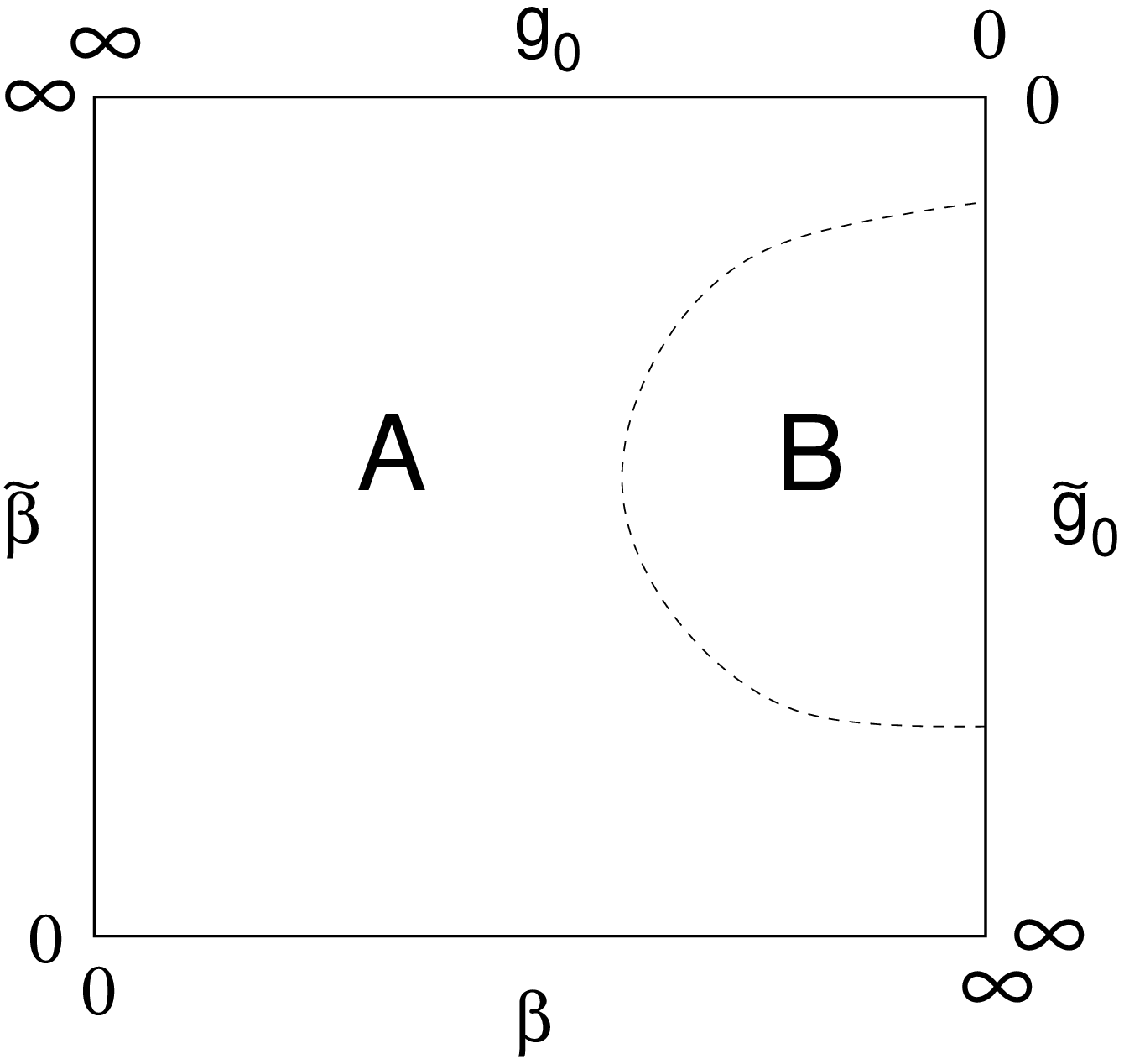}
\hspace{1cm}
\includegraphics*[width=7.0cm]{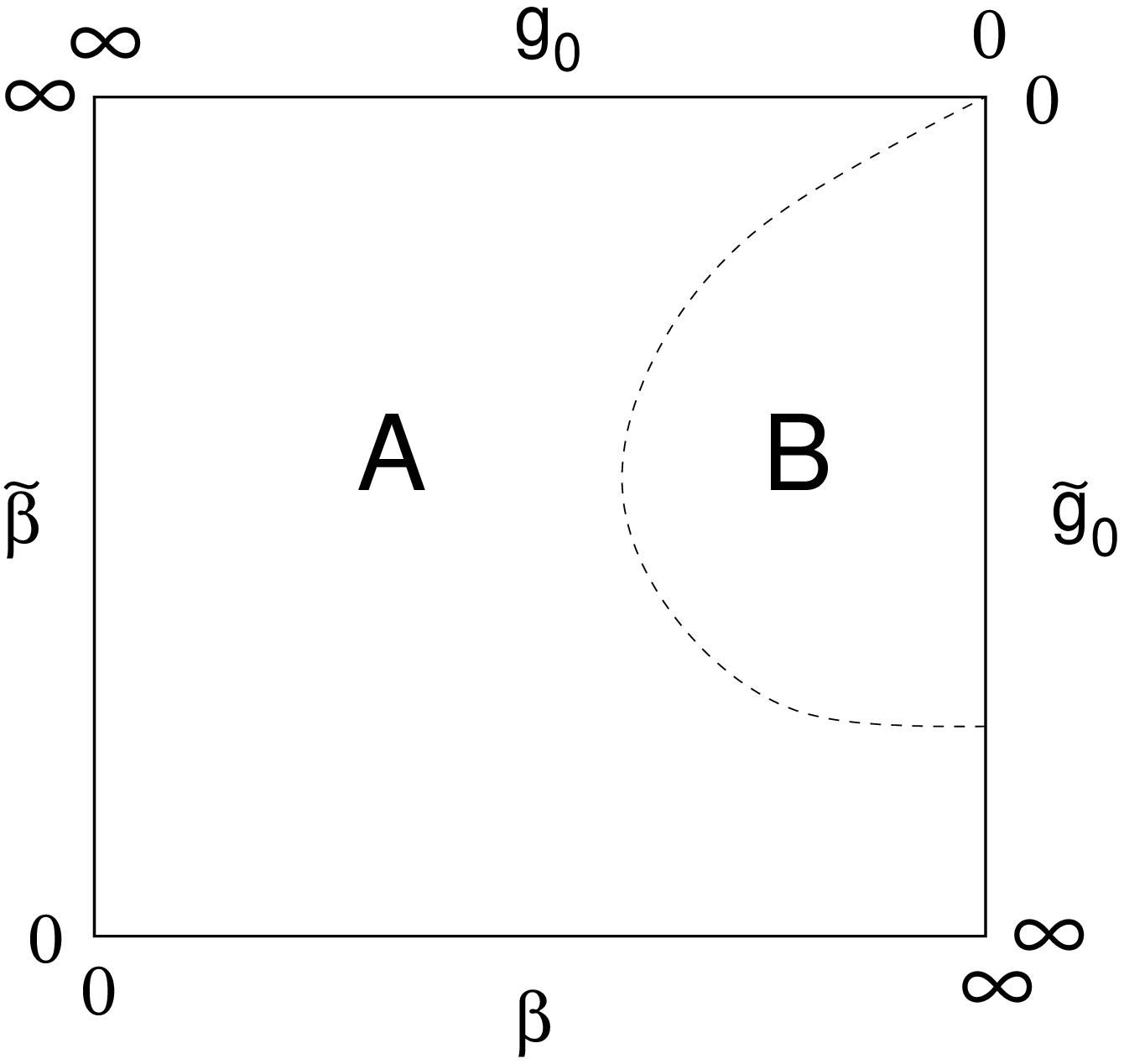}
\vspace*{-2ex}
\begin{quotation}
\floatcaption{phasediag}{%
Two scenarios for the lattice phase diagram \cite{GSMF}.  The axes are
the bare couplings $g_0$ and $\tg_0$, or, equivalently,
$\beta=N/g_0^2$ and $\tilde\beta_0=1/\tg_0^2$.
The confining phase A has a mass gap,
while the Higgs--Coulomb phase B has a massless photon.
Left panel: the Coulomb phase ends at some non-zero $\tg_0$ for $g_0\to 0$.
Right panel: the Coulomb phase extends to the critical point
at $(g_0,\tg_0)=(0,0)$.}
\end{quotation}
\vspace*{-3ex}
\end{figure}

While the scenario of Ref.~\cite{GSMF} covered
the main features of the novel phase and the dynamics that could drive it,
its existence has yet to be established.  Ultimately, this can only be done
via numerical simulations.  But answers to some interesting questions,
even if partial, might come from analytic work.

A particularly important question is the following.
Suppose that the novel phase exists.
In order to be of any relevance for continuum physics, this phase must
extend to the gaussian fixed point at $(g_0,\tg_0)=(0,0)$,
where a continuum limit can be taken.  This scenario is shown in the right
panel in Fig.~\ref{phasediag}, taken from Ref.~\cite{GSMF}.  An alternative
scenario, in which the novel phase is purely a lattice phenomenon
with no bearing on continuum physics, is shown in the left panel.

Some support for the scenario shown in the right panel comes from
the solutions of the one-loop beta function \cite{GSb}.
We have described above a solution where $\tL\gg\L$.
This solution naturally corresponds to approaching
the gaussian fixed point from inside the Higgs--Coulomb phase B,
where the condition $\tg_0/g_0\gg 1$ can be satisfied.
The alternative solution has $\tL\sim\L$ \cite{GSb}.
(There is no solution with $\tL\ll\L$.)
In this case, the non-perturbative dynamics is likely driven by the transverse
sector, with the longitudinal sector playing a passive role.
Therefore it is plausible that that solution corresponds to approaching
the gaussian fixed point from inside the confinement phase A.
In summary, the one-loop beta function is nicely consistent
with the scenario described by the right panel.

However, one has to be careful.  Consider, for instance, the two-loop beta function
of the reduced model.  The two-loop coefficient is not known, and a logical
possibility is that its sign is opposite to that of the one-loop term.
This would suggest the existence of an infrared-attractive fixed point,
in which case no dimensional transmutation would take place
in the continuum limit of the reduced model.
The same situation can alternatively
be described in terms of renormalization-group transformations acting on
the infinite-dimensional space of all marginal and irrelevant operators
that can occur in the reduced model's action:
The flow starting at the gaussian fixed point of the lattice theory
would converge to some $\tg_0^*>0$ when projected back onto the
$g_0=0$ boundary defining the reduced model.
Moreover, we would expect the basin of attraction to extend
on both sides of the fixed point at $\tg_0^*$.  Since phase B
is characterized by its own dynamical scale $\tL$,
it must then be located at stronger values of $\tg_0$ that lie beyond
the basin of attraction.  Phase B would then
be detached from the gaussian fixed point,
which corresponds to the scenario shown on the left panel.

We should note that, even if the two-loop coefficient was known,
this by itself cannot confirm or rule out an infrared-attractive fixed point.
The obvious reason is that a two-loop zero of the beta function
necessarily occurs at a scale where different orders in perturbation theory
(one-loop and two-loop) have become comparable.
But, in general, if different orders in perturbation theory compete,
this means that we went outside the range of validity of perturbation theory.
Once again, ultimately, reliable answers must come from systematic
non-perturbative calculations.

In this paper we will attempt to clarify some of these issues
by appealing to analytic large-$N$ techniques.  We will study
the reduced model for a family of theories where a gauge symmetry $G=SU(N)$
is equivariantly gauge fixed to $H=SU(N-1)\times U(1)$.
The case $N=2$ corresponds to the $SU(2)/U(1)$ theory studied in Ref.~\cite{GSMF},
and the scenarios and questions discussed above generalize to $N\ge2$.

Because the dimensionality of the coset $G/H$ is $2(N-1)$, the number of degrees
of freedom of the reduced model grows only linearly with $N$.  Past experience suggests that models
exhibiting a linear growth can be solved in the large-$N$ limit.%
\footnote{
  This is to be contrasted with $SU(N)$ gauge theories,
  where the number of degrees of freedom grows like $N^2$.
}
A particularly relevant example is the two-dimensional $CP(N)$ model
\cite{DDL1}.  Its large-$N$ solution establishes the generation
of an infrared scale through dimensional transmutation.
In addition, one can infer the infrared particle spectrum from
the large-$N$ solution.

Ideally, a large-$N$ solution of the reduced model would provide
similar information.  Apart from establishing dimensional transmutation,
the knowledge of the infrared spectrum would tell us how the symmetries
of the model are realized, and whether any of them are broken spontaneously.%
\footnote{
  In two dimensions this question is largely moot, since continuous symmetries
  cannot be broken spontaneously.
}
Specifically, if the global $SU(N)$ symmetry of the reduced model
is broken spontaneously to $SU(N-1)\times U(1)$,
this would give rise to $2(N-1)$ NGBs.
As already noted, when the transverse gauge fields
would be turned on, these NGBs would be eaten by the coset gauge fields,
which, in turn, would become massive.
For $N\ge3$, the unbroken $SU(N-1)$ group
would eventually become strong and confining, while the massless photon
associated with the unbroken $U(1)$ will be present for any $N\ge2$.
This is the essence of the Higgs--Coulomb phase B.

The hope is that other symmetries, including the eBRST symmetry itself,
as well as the ghost number symmetry, are not spontaneously broken.  There
are general considerations which indicate that this is a necessary
condition for the existence of a unitary subspace \cite{KO}.

This paper is organized as follows.  In Sec.~\ref{basic} we briefly review
equivariant gauge fixing for the case that $H$ is a maximal subgroup
of $G=SU(N)$.  We describe the reduced model and list its symmetries.
In Sec.~\ref{largeN} we cast the reduced model in a form suitable
for a large-$N$ analysis.  The field $\phi(x)$ takes values in $SU(N)$,
and while, nominally, the number of degrees of freedom grows like $N^2$,
there is a large redundancy due to the local invariance
under $H=SU(N-1)\times U(1)$.
The reformulation of Sec.~\ref{largeN} trades $\phi(x)$ with a unit-norm complex
vector of length $N$.  We furthermore introduce a coset gauge in which
the local $SU(N-1)\times U(1)$ symmetry is eliminated entirely, resulting in one
additional constraint on one of the components of the length-$N$ vector,
consistent with the dimensionality of the coset $G/H$, which is $2(N-1)$.

The gap equation is studied in Sec.~\ref{gap}, and, first,
within a formal large-$N$ approximation,
we find that dimensional transmutation takes place.
Unfortunately, it turns out that the large-$N$ setup of our model
is significantly more complicated than that of the $CP(N)$ model.
A more careful analysis of the large-$N$ limit
confirms the dynamical generation of an infrared scale.
But we are unable to calculate the infrared spectrum,
or to determine the pattern of spontaneous symmetry breaking.
We summarize the progress made in this paper in Sec.~\ref{conclusion}.
Some technical aspects of the eBRST transformation
within the large-$N$ framework are relegated to App.~\ref{s'},
while App.~\ref{MAG} contains a comparison to the body of work using
the maximal abelian gauge.

In this paper we will use continuum notation, but appeal to the lattice
as a non-perturbative regulator when relevant.

\section{\label{basic} Equivariant BRST basics}
We start with a brief review of equivariant gauge fixing for the special case
that a gauge group $G=SU(N)$ is gauge fixed to a maximal subgroup $H$.
We refer to Refs.~\cite{nachgt,FF} for a general discussion.
The maximal subgroup is defined as the subgroup whose generators commute with
the diagonal matrix%
\footnote{Notice that $\tT_0$ squares to the identity matrix.
  For $N>2$, it is a linear combination of the generators
  of the Cartan subgroup and of the identity matrix itself.
}
\begin{equation}
  \tT_0 = {\rm diag}(\, \underbrace{+1,+1,\cdots,+1}_{\mbox{$N-M$ times}}\,, \
                    \underbrace{-1,-1,\cdots,-1}_{\mbox{$M$ times}}\, ) \ .
\label{ttilde}
\end{equation}
For $1<M\le [N/2]$, the maximal subgroup is $H=SU(N-M)\times SU(M) \times U(1)$,
whereas for $M=1$ it is $SU(N-1)\times U(1)$.
We split the gauge field $V_\m\in SU(N)$ as
\begin{subequations}
  \label{V}
\begin{equation}
  V_\m = A_\m + W_\m \ ,
\label{Va}
\end{equation}
where
\begin{equation}
  A_\m = \half ( V_\m + \tT_0 V_\m \tT_0)\ ,
\label{Vb}
\end{equation}
is the $H$ gauge field whose generators commute with $\tT_0$, while
\begin{equation}
  W_\m = \half ( V_\m - \tT_0 V_\m \tT_0)\ ,
\label{Vc}
\end{equation}
\end{subequations}
contains the $G/H$ coset generators
(these generators anti-commute with $\tT_0$).
Like $W_\m$, the ghost fields $C$ and $\bc$ live in the coset,
and are (formally) hermitian.

The on-shell gauge-fixing lagrangian is
\begin{equation}
  \cl = \cl_\phi + \cl_{gh} \ ,
\label{Lon}
\end{equation}
where the bosonic part is
\begin{equation}
  \cl_\phi =
  \frac{1}{\tg^2}\,\tr\!\left((D_\m(A) W_\m)^2\right) \ ,
\label{Lphi}
\end{equation}
and the ghost part is
\begin{equation}
  \cl_{gh} =
  -2\, \tr\!\left(\bc D_\m(A) D_\m(A) C \right)
  + 2\,\tr\!\left( [W_\m,\bc]\, [W_\m,C] \right)
  -\tg^2\,\tr(\tX^2) \ .
\label{Lgh}
\end{equation}
Here
\begin{equation}
  \tX = i \{ C,\bc \} \ ,
\label{Xtilde}
\end{equation}
and $D_\m(A)$ is the $H$-covariant derivative, \eg,
\begin{equation}
  D_\m(A)C = \partial_\m C + i[A_\m, C] \ .
\label{DA}
\end{equation}
We note that the four-ghost coupling in Eq.~(\ref{Lgh}) is the key reason that
the ``no-go'' theorem of Ref.~\cite{HNnogo} is avoided.

The equivariantly gauge-fixed theory is invariant under local $H$
transformations.  The on-shell eBRST transformations are%
\footnote{
  There is also an anti-eBRST symmetry in which the roles of the $C$ and
  $\bc$ ghost fields are flipped, see Ref.~\cite{nachgt}.
}
\begin{subequations}
\label{brst}
\begin{eqnarray}
  sV_\m &=& \partial_\m C +i[V_\m,C]\ ,
\label{brsta}\\
  sC &=& 0 \ ,
\label{brstb}\\
  s\bc &=& \frac{1}{\tg^2}\, D_\m(A)W_\m\ .
\label{brstc}
\end{eqnarray}
\end{subequations}
The ghost-number symmetry extends to an $SL(2,R)$ symmetry \cite{MS1}.
This symmetry has the same algebra as $SU(2)$, and the two other generators
act as raising or lowering operators that change the ghost number by $\pm2$.

We next turn to the reduced model, which is obtained
by constraining the gauge field to the trivial orbit,
\begin{equation}
  V_\m = -i \phi \partial_\m \phi^\dagger \ .
\label{Vphi}
\end{equation}
The eBRST transformation rule of the gauge field, Eq.~(\ref{brsta}), is replaced by
\begin{equation}
  s\phi = -iC \phi \ .
\label{sphi}
\end{equation}
Apart from the substitution~(\ref{Vphi}), the on-shell lagrangian
as well as the eBRST transformation rules of the ghost fields are unchanged.
The reduced model inherits the $SL(2,R)$ symmetry.
In addition, it is invariant under a local $H$ symmetry
and a global $G$ symmetry, which act on the $\phi$ field according to
\begin{equation}
  \phi(x) \to h(x)\phi(x) g^\dagger\ ,
\label{phitrans}
\end{equation}
with $h(x) \in H$, and $g\in G$.  The ghost fields are singlets
under the global $G$ symmetry, and transform under the local $H$
symmetry according to
\begin{equation}
  C(x) \to h(x) C(x) h^\dagger(x) \ , \qquad
  \bc(x) \to h(x) \bc(x) h^\dagger(x) \ .
\label{ghtrans}
\end{equation}
The original equivariantly gauge fixed theory can be reconstructed
from the reduced model by promoting the global $G$ symmetry
to a local symmetry.%
\footnote{
  This gives rise to the so-called Higgs picture of the original theory,
  in which both $V_\m$ and $\phi$ are present as independent fields, and
  which has separate $H$ and $G$ local symmetries,
  see Refs.~\cite{nachgt,GSMF}.
}

In the rest of this paper we will limit the discussion to $M=1$,
namely, to the maximal subgroup $H=SU(N-1)\times U(1)$.
The diagonal matrix $\tT_0$ of Eq.~(\ref{ttilde}) takes the explicit form
\begin{equation}
  \tT_0 = {\rm diag}(\,\underbrace{+1,+1,\cdots,+1}_{\mbox{$N-1$ times}}\,, -1 )\ .
\label{ttil}
\end{equation}

\section{\label{largeN} Large-$N$ action}
In order to facilitate a large-$N$ treatment, we will replace
the unitary field $\phi$ by a more economic representation of the
coset degrees of freedom.  In Sec.~\ref{boson} we start with the bosonic part
of the action,%
\footnote{
  We use the notion of a boson field in a loose sense, referring to the fact
  that its elements are $c$-numbers, and not Grassmann-numbers.
}
showing that it can be re-expressed in terms of a unit-norm
complex vector of length $N$, denoted by $z$.
The $SU(N-1)$ part of the local $H$ symmetry acts trivially on $z$.
The $U(1)$ factor acts non-trivially, effectively eliminating one
degree of freedom.  Another degree of freedom is eliminated by the
norm constraint, so that the true number of degrees of freedom
is $2(N-1)$, in agreement with the dimensionality of the coset.

In Sec.~\ref{coset} we introduce a coset gauge, in which a unique representative
is selected for each element of the coset, and the local invariance
including the $U(1)$ factor is eliminated altogether.
This representation involves
$2(N-1)$ unconstrained real fields.
We then use the coset gauge in Sec.~\ref{ghost}
to extract the part of the ghost action which is needed
for the large-$N$ saddle point equations that we will study in Sec.~\ref{gap}.
A brief summary in Sec.~\ref{smry} highlights some technical difficulties
that, in the next section, will turn out to significantly limit the
scope of our large-$N$ treatment.

\subsection{\label{boson} Boson sector}
We begin by writing the unitary field $\phi\in SU(N)$ in block form
\begin{equation}
\phi = \left(\begin{array}{cc}
         R & \vn \\
         \vm^\dagger & c
       \end{array}\right)\ ,
\label{whi}
\end{equation}
in which $R$ is an $(N-1)\times (N-1)$ matrix, $\vn$ and $\vm$ are vectors
of length $N-1$, and $c$ is a complex number.
The blocks containing $R$ and $c$ commute with $\tT_0$ of Eq.~(\ref{ttil}),
whereas those containing $\vn$ and $\vm$ anti-commute.
Unitarity of the matrix $\phi$ implies the relations
\begin{eqnarray}
  R^\dagger R+\vm\vm^\dagger &=& I_{N-1} \ ,
\label{unirels}\\
  \vn^\dagger\vn+c^*c = \vm^\dagger\vm+cc^* &=& 1 \ ,
\nonumber\\
  c\vm + R^\dagger\vn &=& 0\ ,
\nonumber
\end{eqnarray}
where $I_K$ is the $K\times K$ identity matrix.
We introduce a length-$N$ complex vector,
\begin{equation}
  z = \left(\begin{array}{c}
        \vm \\
        c^*
      \end{array}\right) \ ,
\label{z}
\end{equation}
which, according to Eq.~(\ref{unirels}), has a unit norm,
\begin{equation}
\label{znorm}
z^\dagger z=1\ .
\end{equation}

We now turn to the bosonic part of the lagrangian, Eq.~(\ref{Lphi}).
We start with the relation
\begin{equation}
  i \phi^\dagger\, D_\m(A) W_\m \,\phi
  =  (\bo\cp) \cp -  \cp (\bo\cp) \ ,
\label{wDW}
\end{equation}
with the $N\times N$ projection operator
\begin{equation}
  \cp = z  z^\dagger \,.
\label{Pz}
\end{equation}
It is straightforward to check Eq.~(\ref{wDW}) by substituting Eq.~(\ref{Vphi})
into Eq.~(\ref{V}) and using this on the left-hand side, while
expressing $\cp$ on the right-hand side in terms of the operator $\tA$
introduced in App.~\ref{s'}.
Substituting Eq.~(\ref{wDW}) into Eq.~(\ref{Lphi}) gives
\begin{equation}
  \cl_\phi
  =  \frac{2}{\tg^2}\, z^\dagger (\bo\cp) (1-\cp)(\bo\cp) z \ .
\label{DWsq}
\end{equation}
We comment that since $1-\cp$, too, is a projection operator,
this result is non-negative, as it should.

Using Eqs.~(\ref{phitrans}) and~(\ref{whi}), it can be checked that $z$
transforms in the fundamental representation of the global $SU(N)$ symmetry.
It is inert under the local $SU(N-1)\subset H$, while under the
local $U(1)\subset H$ it transforms according to
\begin{equation}
\label{wab}
z\to e^{-i\th}z\ .
\end{equation}
The projector $\cp$ is invariant under the $U(1)$ symmetry.
Thanks to the constraint~(\ref{znorm}), the vector field
\begin{equation}
  a_\m
  = \frac{i}{2} (z^\dagger \partial_\m z - \partial_\m z^\dagger z) \ ,
\label{amu}
\end{equation}
transforms as an abelian gauge field,
\begin{equation}
  a_\m \to a_\m + \partial_\m \th \ ,
\label{vtrans}
\end{equation}
and the corresponding covariant derivative is
\begin{equation}
  D_\m z = \partial_\m z +ia_\m z \ .
\label{Da}
\end{equation}
Equations~(\ref{znorm}) and~(\ref{amu}) also imply
\begin{equation}
  z^\dagger D_\m z = z^\dagger (\partial_\m +ia_\m) z = 0 \ .
\label{vconst}
\end{equation}

The bosonic lagrangian can be further simplified.  Using that
\begin{equation}
  (1-\cp) (\bo \cp) z = (1-\cp) (D^2 z) \ ,
\label{ImPz}
\end{equation}
we may rewrite
\begin{equation}
  \cl_\phi = \frac{2}{\tg^2}\, (D^2 z)^\dagger (1-\cp) (D^2 z) \ .
\label{simpler}
\end{equation}
In addition,
\begin{equation}
  (D^2 z)^\dagger \cp (D^2 z)
  = |z^\dagger D^2 z|^2
  = ((D_\m z)^\dagger D_\m z)^2 \ .
\label{further}
\end{equation}
Here we have used Eq.~(\ref{vconst}),
which implies that $\partial_\m(z^\dagger D_\m z)=0$.
The bosonic lagrangian can now be written as
\begin{equation}
  \cl_\phi =
  \frac{2}{\tg^2}\, (D^2 z)^\dagger (D^2 z)
  + \frac{\tg^2}{2}\, \s^2 + 2\s (D_\m z)^\dagger D_\m z
  + i\a(z^\dagger z - 1) + \z_\m\, z^\dagger D_\m z\ ,
\label{quadS}
\end{equation}
in which $z$ is an unconstrained complex $N$-vector.
We have traded the rightmost expression in Eq.~(\ref{further}) with the terms
that depend on the auxiliary field $\s$.
Following the usual treatment of the $CP(N)$ model \cite{DDL1},
the constraint~(\ref{znorm}) is enforced via a Lagrange multiplier $\a$.
The gauge field $a_\m$ is treated as an independent field,
and the constraint~(\ref{vconst}) is enforced by a vector
Lagrange multiplier $\z_\m$.

Equation~(\ref{quadS}) is amenable to large-$N$ methods because
it is bilinear in the length-$N$ vector $z$, which is now unconstrained.
Large-$N$ counting becomes manifest if we introduce the `t~Hooft
coupling $\tl=\tg^2 N$, and perform the rescalings $z\to \tg z$
and $\s\to \s/\tg^2$, along with similar rescalings
for the auxiliary fields $\a$ and $\z_\m$.
The final form of the bosonic lagrangian is
\begin{equation}
  \cl_\phi =
  2 (D^2 z)^\dagger (D^2 z)
  + \frac{N}{2\tl}\, \s^2 + 2\s (D_\m z)^\dagger D_\m z
  + i\a(z^\dagger z - N/\tl) + \z_\m\, z^\dagger D_\m z\ ,
\label{Srescale}
\end{equation}
where now every single degree of freedom is treated as $O(1)$ in the large-$N$
counting.

As we will soon see, things are more complicated in the ghost sector.
But, before dealing with the ghost sector, we first introduce the coset gauge.

\subsection{\label{coset} Coset gauge}
The local $H$ invariance of the reduced model
means that the field $\phi$ really takes values
in the coset $G/H$.  In loose analogy with the unitary gauge of ordinary
gauge theories, we will now define a coset gauge, which selects
a unique representative for each coset element.  To this end we write
\begin{equation}
  \phi = \phi_H\,\phi_{G/H}\ ,
\label{phiHGH}
\end{equation}
where $\phi_H\in H$, and
\begin{eqnarray}
\phi_{G/H}&=&
  \exp\left[i\left(\begin{array}{cc}
    0 & \vv \\
    \vv^\dagger & 0
  \end{array}\right)\right]
  \ = \ \left(\begin{array}{cc}
    1+(\cos{v}-1)\hv\hv^\dagger & i\hv\sin{v} \\
    i\hv^\dagger\sin{v} & \cos{v}
  \end{array}\right) \ ,
\label{phiGH}\\
\rule{0ex}{4ex}
v &=& \sqrt{\vv^\dagger\vv}\ , \qquad \hv \ = \ \vv/v\ ,
\nonumber
\end{eqnarray}
with $\vv$ an $(N-1)$-dimensional complex vector.
We may now completely fix the $H$-invariance by setting $\phi_H$ equal
to the identity matrix.  Every coset element is now represented as
$\phi = \phi_{G/H}$.  This representation involves the $2(N-1)$
real components of $\vv$, which is the correct number of degrees of freedom.
Comparing Eq.~(\ref{phiGH}) with Eq.~(\ref{whi}) gives rise to
\begin{eqnarray}
\label{wcomp}
c&=&\cos{v}\ ,\\
\vn&=&i\hv\sin{v}\ ,\nonumber\\
\vm&=&-i\hv\sin{v}\ ,\nonumber\\
R&=&1+(\cos{v}-1)\hv\hv^\dagger=R^\dagger\ ,
\nonumber
\end{eqnarray}
and thus
\begin{equation}
\label{phiGHmc}
  \phi_{G/H} = \left( \begin{array}{cc}
    1-\frac{\vec{m}\vec{m}^\dagger}{1+c} & -\vec{m} \\
    \vec{m}^\dagger & c
  \end{array}\right) \ .
\end{equation}
The nominal number of degrees of freedom in the vector $z$ of Eq.~(\ref{z}) is $2N$.
We can see from Eq.~(\ref{wcomp}) how two degrees of freedom are removed.
First, the norm constraint~(\ref{znorm}) is automatically satisfied.
Moreover, the $N$-th component, $c$, is now real.%
\footnote{
  The phase of the $U(1)$ transformation~(\ref{wab}) may, thus,
  be identified with the phase of $c$.
}

Of the original continuous symmetries in Eq.~(\ref{phitrans}),
the only surviving symmetry in the coset gauge is the diagonal,
global $SU(N-1)$ subgroup, under which
\begin{equation}
  \phi_{G/H}(x) \to g\, \phi_{G/H}(x)\, g^\dagger \ , \qquad g \in SU(N-1) \ ,
\label{diagGH}
\end{equation}
which corresponds to $\vec{v}\to g\vec{v}$, and
with similar transformation rules for the ghost fields.

The eBRST transformation rules have to be modified in order to preserve
the coset gauge.  The new transformation rules are worked out in App.~\ref{s'}.

\subsection{\label{ghost} Ghost sector}
Our goal is to express the complete action using a set of degrees of freedom
that grows linearly with $N$.  As for the ghost fields themselves, we have
\begin{equation}
\label{ghrew}
  C = \half \left(\begin{array}{cc}
               0 & \vc \\
               \vc^\dagger & 0
            \end{array}\right)
    = \half \left(\begin{array}{cc}
               0 & \vc_1-i\vc_2 \\
               \vc_1+i\vc_2 & 0
            \end{array}\right)\ ,
\end{equation}
where $\vc_1$ and $\vc_2$ are $(N-1)$-dimensional vectors with Grassmann
elements.  There is a similar expression for $\bc$.
The number of ghosts degrees of freedom thus grows linearly with $N$.
The ghost action depends on the $\phi$ field, too, and, in order
to express this dependence using order-$N$ degrees of freedom, we will
resort to the coset gauge.

The result, obtained by substituting Eq.~(\ref{Vphi}) with $\phi=\phi_{G/H}$
into Eq.~(\ref{Lgh}), is quite complicated.
In the rest of this subsection, we will work out those parts
of the ghost action that are leading in large-$N$ counting.

We begin by performing the same rescaling as we did in the transition
from Eq.~(\ref{quadS}) to~(\ref{Srescale}), namely,
$\vm\to \tg\vm$, $c\to \tg c$.  We see that any $c$ dependence will
necessarily be through the product $\tg c$.  Since the coupling $\tg$
is parametrically of order $1/\sqrt{N}$,
we may neglect the $c$ dependence everywhere.
By contrast, the vector $\vm$ has $N-1$ components, and so, for example,
when $\vm^\dagger$ is contracted with $\vm$ in the
operator $\tg^2 \vm^\dagger\cdt\vm$, this will produce a contribution
proportional to $\tg^2(N-1)\approx \tl$.  Therefore, this operator
is $O(1)$ in the large-$N$ counting,
and the same is true in general for $\tg\vm$.

Equation~(\ref{phiGHmc}) now simplifies to
\begin{equation}
  \f_{G/H} = \left( \begin{array}{cc}
    1-\tg^2\vec{m}\vec{m}^\dagger & -\tg\vm \\
    \tg\vm^\dagger & 0
  \end{array}\right) + O(1/\sqrt{N}) \ ,
\label{phim}
\end{equation}
and substituting this into Eq.~(\ref{Vphi}) gives
\begin{equation}
  iV_\m =
  \left( \begin{array}{cc}
    - \tg^2 (\partial_\m \vm) \vm^\dagger
    + \tg^2 \vm (\partial_\m \vm)^\dagger
    & \tg \partial_\m \vm \\
    -\tg \partial_\m \vm^\dagger & 0
  \end{array}\right) + O(1/\sqrt{N}) \ .
\label{Vmuphi}
\end{equation}
Substituting Eq.~(\ref{Vmuphi}) into Eq.~(\ref{V}),
and using this for the bilinear terms in the ghost action~(\ref{Lgh}), yields
\begin{subequations}
\label{WDC}
\begin{eqnarray}
  2\,\tr\!\left( [W_\m,\bc]\, [W_\m,C] \right)
  &=& -\tg^2\, \partial_\m \vm^\dagger \cdt \partial_\m \vm\; \vbc_{\a}\cdt\vc_{\a}
  + O(1/\sqrt{N})\ ,
\hspace{5ex}
\label{WC}\\
  2\, \tr\Big((D_\m(A) \bc)  (D_\m(A) C)\Big)
  &=& \partial_\m \vbc_{\a} \cdt \partial_\m \vc_{\a} + O(1/\sqrt{N}) \ .
\label{DC}
\end{eqnarray}
\end{subequations}
Summation over $\a=1,2,$ is implied (see Eq.~(\ref{ghrew})).
The dot notation indicates
the inner product between two $(N-1)$-dimensional vectors.
The only contractions we have kept in Eq.~(\ref{WDC}) are those that
can produce an additional factor of $N$ in diagrams.
The terms we have neglected include
mixed boson-ghost double contractions such as, for example,
$\tg^2 \partial_\m \vm^\dagger \cdt \vbc \, \vc^\dagger \cdt \partial_\m \vm$.
We have also dropped any dependence on the abelian gauge field~(\ref{amu}).
By Lorentz invariance,
$\langle z^\dagger \partial_\m z \rangle =
\langle \partial_\m z^\dagger  z \rangle =0$, and, as a result,
any term that depends on the $a_\m$ gauge field cannot contribute
to the large-$N$ limit.

It remains to deal with the four-ghost term in Eq.~(\ref{Lgh}), which,
in preparation for the next step, we rewrite as%
\footnote{
  On the right-hand side of Eq.~(\ref{S4noaux}), the first term and the
  last two terms are separately invariant under $SL(2,R)$.
}
\begin{equation}
  - \tr (\tX^2)
  = - \frac{1}{8} (\vbc_{\a}\cdt\vc_{\a})^2
    + \, \frac{1}{2} \vbc_{1}\cdt\vbc_{2}\, \vc_{1}\cdt\vc_{2}
    - \frac{1}{8} (\e_{\a\b}\, \vbc_{\a} \cdt \vc_{\b})^2 \ .
\label{S4noaux}
\end{equation}
Let us put together the boson and ghost terms.
First we eliminate the auxiliary field $\s$ of Eq.~(\ref{Srescale})
(soon to be replaced by a different auxiliary field),
and then drop any terms that do not contribute to the large-$N$ limit.
This gives
\begin{equation}
  \cl_\phi =
  2\, \bo\vm^\dagger\cdt\bo\vm
  + 2\frac{\tl}{N}\, (\partial_\m \vm^\dagger \cdt \partial_\m \vm)^2
  + i\a(\vm^\dagger \cdt \vm - N/\tl) + O(1/\sqrt{N}) \ .
\label{SleadN}
\end{equation}
As we did with the ghost action, we dropped the dependence on the $c$ field
and on the auxiliary gauge field $a_\m$, which belongs
in the $O(1/\sqrt{N})$ part.  (Therefore we dropped the term
that enforces the constraint~(\ref{vconst}), too.)
Using also Eqs.~(\ref{WDC}) and~(\ref{S4noaux}) we obtain
\begin{eqnarray}
  \cl &=& 2\,\bo\vm^\dagger\cdt\bo\vm +\partial_\m\vbc_{\a}\cdt\partial_\m \vc_{\a}
\label{SN}\\
  &&
  +\frac{N}{2\tl}\,\eta^2
  +2\eta\left(\partial_\m \vm^\dagger \cdt \partial_\m \vm
  +\frac{1}{4}\,\vbc_{\a}\cdt\vc_{\a}\right)
\NON
  &&
  +\frac{N}{\tl}\left(\frac{\r_0^2}{2}+\r^*\r\right)
  +\frac{\r_0}{2}\, \e_{\a\b}\, \vbc_{\a} \cdt \vc_{\b}
  +\frac{1}{\sqrt{2}}\left(\r\, \vbc_{1}\cdt\vbc_{2}
  -\r^* \vc_{1}\cdt\vc_{2} \right)
\NON
  &&
  + \rule{0ex}{3ex}
  i\a\left(\vm^\dagger \cdt \vm -N/\tl\right)
  + O(1/\sqrt{N}) \ .
\nonumber
\end{eqnarray}
By introducing the new auxiliary field $\eta$,
we take advantage of the fact that
the contributions from the second term on the right-hand side
of Eq.~(\ref{SleadN}), from Eq.~(\ref{WC}),
and from the first term on the right-hand side of Eq.~(\ref{S4noaux}),
together form a perfect square.%
\footnote{
  The $\h$ dependence is only through the terms
  shown explicitly on the right-hand side of Eq.~(\ref{SN}).
  By construction, the $O(1/\sqrt{N})$ terms do not depend on $\h$.
  Note that this auxiliary field is similar to, but not the same as the
  field $\s$ in Eq.~(\ref{quadS}).
}
The third line of Eq.~(\ref{SN}) is obtained by
applying similar Hubbard--Stratonovich transformations
to the last two terms in Eq.~(\ref{S4noaux}).

Finally, we note that neglecting the $c$ dependence everywhere
requires a slightly more careful consideration.  To explain this point,
consider a vertex with the generic form $(\tg c)^n \bc C$.
Contracting the ghost fields $C$ and $\bc$ into a loop
can produce a factor of $N$, and thus this vertex is of order $N^{1-n/2}$.
The dangerous case is $n=1$, for which the vertex is of order $\sqrt{N}$.
We have verified the absence of such $n=1$
vertices in the ghost action by explicit calculation.

\subsection{\label{smry} Summary}
For the purely bosonic action we found in Sec.~\ref{boson} that,
using the unitarity relations~(\ref{unirels}), it can be expressed in terms
of the $N$-vector $z$, as can be seen in Eqs.~(\ref{DWsq}) or~(\ref{quadS}).
The matrix $R$ of Eq.~(\ref{whi}), whose number of elements grows like $N^2$,
drops out.

Unlike the bosonic action, we have not been able to cast
the ghost action in a form that will not depend on $R$, if
the parametrization~(\ref{whi}) is used.  In order to make progress
we had to resort to the coset gauge.  The resulting lagrangian, Eq.~(\ref{SN}),
is indeed bilinear in all the fields whose size grows linearly with $N$,
making it amenable to a large-$N$ treatment.

However, this now comes at a new price.
The large-$N$ lagrangian~(\ref{SN}) has been obtained from the original
lagrangian~(\ref{Lon}) by neglecting $O(1/\sqrt{N})$ terms.
Superficially, the theories defined by the lagrangians~(\ref{Lon})
and~(\ref{SN}) ought to have the same large-$N$ limit.
As we will see in the next section, the actual situation is
more subtle, and this will significantly limit our ability to solve
the theory by applying large-$N$ methods.

\section{\label{gap} Gap equation and dimensional transmutation}
In this section we proceed to study the large-$N$ solution
of the reduced model.  In Sec.~\ref{gapN} we begin by working out
the saddle-point equations while dropping all $O(1/\sqrt{N})$ terms
in the lagrangian~(\ref{SN}).  Consistent with the asymptotically free
one-loop beta function, we find that the ground state exhibits dimensional
transmutation, and that an infrared scale $\tL$ is generated dynamically.
We also discuss some consistency checks of the solution.

In Sec.~\ref{erole} we point out that, at face value, our large-$N$ solution
implies that eBRST symmetry is broken spontaneously.  If this were true, this
would be bad news, because it is generally believed that an unbroken
BRST-type symmetry is a necessary condition for unitarity \cite{KO}.

In Sec.~\ref{limitation} we return to the terms we have dropped from the lagrangian
in our large-$N$ solution, and explain why it was in fact unjustified
to neglect them at low energies.  The effects of these terms are controlled
by the running coupling, and, when the latter becomes strong
at the scale $\tL$, the resulting contributions to physical
observables are no longer suppressed by a small parameter
(in spite of the formal suppression by $1/\sqrt{N}$).
This is true, in particular, for the two-point functions of the
scalar field $\vm$ and the ghost fields, which
determine the large-$N$ gap equation.
Fortunately, thanks to the logarithmic enhancement of the contribution
from scales $p\gg\tL$, the conclusion we found in Sec.~\ref{gapN},
that dimensional transmutation takes place, remains valid.

In other words, while an infrared scale $\tL$ is indeed generated
dynamically for large $N$, the $1/N$ expansion is not tractable beyond this
conclusion,
and does not allow us to determine the
non-perturbative physics at that scale.
In particular, the lack of sufficient knowledge of $p\sim\tL$ physics means that
it is not possible to use our large-$N$ solution to determine
how the symmetries of the model are realized.
In Sec.~\ref{ir} we elaborate on the relevant questions, but answering
them goes beyond the scope of this paper, and may ultimately have to come
from numerical simulations.

\subsection{\label{gapN} Gap equation}
We begin with a study of the solution of the theory in the large-$N$ limit,
dropping the additional terms indicated as $O(1/\sqrt{N})$ corrections
in Eq.~(\ref{SN}).  Later on we will re-examine the validity of this procedure.

We will employ continuum notation, assuming an ultraviolet cutoff $M$.
Ultimately, one has to use the lattice as a non-perturbative regulator.
With this in mind, we will treat Eq.~(\ref{SN}) as the bare
lagrangian, replacing $\tg$ with the bare (lattice) coupling $\tg_0$,
and $\tl$ with $\tl_0=\tg_0^2 N$.
In solving the gap equations below, we may self-consistently neglect
discretization effects, because they are suppressed
relative to the terms we keep.

We find it convenient to first ignore the norm constraint~(\ref{znorm}),
and the corresponding Lagrange multiplier $\a$.  Later we will verify
that our solution is consistent with the norm constraint.
We will study the effective potential as a function of the
vacuum expectation values $\h=\svev{\h(x)}$ and $\r_0=\svev{\r_0(x)}$.%
\footnote{
  Thanks to $SL(2,R)$ invariance, the effective potential must be a function
  of $\r_0^2/2+\r^*\r$, where $\r=\svev{\r(x)}$.
  This allows us to set $\r=0$, and study the dependence on $\r_0$ only.
}
Integrating over the scalar
fields $\vm,\vm^\dagger$ and the ghost fields $\vc_{\a},\vbc_{\a}$ we find
\begin{eqnarray}
  v_{eff} \equiv \frac{V_{eff}}{N}
  &=& \frac{\h^2+\r_0^2}{2\tl_0}
\label{veff}\\
  && + \int^M \frac{d^4p}{(2\p)^4} \left(
          \log[2p^2(p^2+\h)] - \log[(p^2+\h/2)^2+\r_0^2/4] \right) \ .
\nonumber
\end{eqnarray}
The first logarithm inside the integral comes from
the scalar degrees of freedom, and the second from the ghosts.
The saddle-point equations are
\begin{subequations}
\label{gapeq}
\begin{eqnarray}
  \frac{\h}{\tl_0} &=& \frac{1}{4} \int^M \frac{d^4p}{(2\p)^4}
  \frac{2p^2\h + \h^2 - \r_0^2}{[(p^2+\h/2)^2+\r_0^2/4](p^2+\h)} \,,
\label{gapeqa}\\
  \frac{\r_0}{\tl_0} &=& \rule{0ex}{3ex} \frac{1}{2} \int^M \frac{d^4p}{(2\p)^4}
  \frac{\r_0}{(p^2+\h/2)^2+\r_0^2/4} \,.
\label{gapeqb}
\end{eqnarray}
\end{subequations}
We will assume $\h\ge0$.
A negative value would give rise to an infrared divergence
in Eq.~(\ref{gapeqa}) because of the factor of $p^2+\h$ in the denominator,
showing that the true vacuum cannot be at negative $\h$.
As for $\r_0$, we may take it to be non-negative without loss of generality.

In both of the saddle-point equations, the integral is dominated by
a logarithmic contribution, that ranges from the ultraviolet cutoff $M$
of the bare (lattice) theory, down to some dynamically generated scale $\tL$,
where $\tL^2=\max(\h,\r_0)$.  Keeping only the logarithmic piece,
either one of the gap equations simplify to%
\footnote{
  Here we are assuming that $\h$ and $\r_0$ are not both zero,
  as we will verify below.
}
\begin{equation}
\label{gaplog}
  \frac{2}{\tl_0} =  \int^M_{\tL} \frac{d^4p}{(2\p)^4} \frac{1}{(p^2)^2} \ ,
\end{equation}
or
\begin{equation}
\label{gapsolv}
  \frac{16\p^2}{\tl_0} = \log(M/\tL) \ ,
\end{equation}
a result which is consistent with the one-loop renormalization-group
equation \cite{GSb}
\begin{equation}
  \frac{\partial \tg^2}{\partial \log\m} = -\frac{N}{16\p^2}\, \tg^4 \ .
\label{betafnred}
\end{equation}
It follows that dimensional transmutation takes place in the large-$N$ limit.
This has an interesting corollary.
Because the large-$N$ solution~(\ref{gapsolv}) is consistent with
vanishing two-loop and higher-loop coefficients in the beta function,
this solution rules out a conformal infrared behavior
for the large-$N$ reduced model.

We next turn to the individual expectation values, $\h$ and $\r_0$.
At face value, the gap equations can have a few qualitatively different
solutions: (1) $\h=\r_0=0$; (2) $\h>0$, $\r_0=0$; and (3) $\h>0$, $\r_0>0$.
A fourth solution where $\r_0>0$ but $\h=0$ is immediately ruled out by
Eq.~(\ref{gapeqa}).
It is also easy to show that the perturbative vacuum $\h=\r_0=0$ is unstable.
To this end we first set $\r_0=0$, so that, from its definition,
$\tL=\sqrt\h$.  Using Eqs.~(\ref{gapeqa}) and~(\ref{gapsolv}) we
obtain the second $\h$-derivative
\begin{equation}
  \frac{\partial^2 v_{eff}}{\partial^2\h} \bigg|_{\r_0=0}
  = \frac{1}{\tl_0} - \frac{\log(M^2/\h)}{32\p^2} + \mbox{constant} \ .
\label{d2eta}
\end{equation}
For $\h\to0$, the curvature tends to $-\infty$.
A similar conclusion applies to the curvature in the $\r_0$ direction.
Therefore, the point $\h=\r_0=0$ is a (singular) local maximum.

A solution where both $\h,\r_0>0$ does not exist either.
Write the numerator of Eq.~(\ref{gapeqa}) as $2\h(p^2+\h)-(\h^2+\r_0^2)$,
and note that for the first term we may cancel a factor $p^2+\h$
between the numerator and the denominator.  Next,
assuming $\r_0\ne0$, we may divide both sides of Eq.~(\ref{gapeqb}) by it, and
then use the result on the right-hand side of Eq.~(\ref{gapeqa}), which becomes
\begin{equation}
  \frac{\h}{\tl_0} - \frac{1}{4} \int^M \frac{d^4p}{(2\p)^4}
  \frac{\h^2 + \r_0^2}{[(p^2+\h/2)^2+\r_0^2/4](p^2+\h)} \,.
\label{simplifygap}
\end{equation}
Since by assumption both $\h$ and $\r_0$ are positive, the right-hand side of
Eq.~(\ref{gapeqa}) is now seen to be strictly smaller than the left-hand side,
which is a contradiction.  Therefore, such a solution is impossible.

The only remaining solution is characterized by $\h>0$ and $\r_0=0$.
Since $v_{eff}$ is bounded below, this solution is the absolute minimum
of the potential.  The dynamically generated scale in Eq.~(\ref{gapsolv})
is therefore
\begin{equation}
  \tL = \sqrt{\eta} \ .
\label{tL}
\end{equation}

We next turn to the norm constraint.  Neglecting $c$, it reads
$\vm^\dagger \vm=\tg_0^{-2}$, where $\vm$ is the rescaled bare field.
Using the solution of the gap equations we have found above,
it is easily seen that
\begin{equation}
  \svev{\vm^\dagger \vm}
  = \frac{N}{2} \int^M \frac{d^4p}{(2\p)^4}\, \frac{1}{p^2(p^2+\tL^2)}
  = \frac{N}{16\p^2} \, \log(M/\tL) = \frac{1}{\tg_0^2} \,,
\label{zdagz}
\end{equation}
where we have applied the same approximations as before.
We conclude that our solution respects the norm constraint, as it should.

Before we turn our attention to the question of how the various symmetries
are realized, there is one more consistency check that we can do.
Equation~(\ref{SN}) contains a term
$2a_1\eta\,\partial_\m \vm^\dagger \cdt \partial_\m \vm$ with $a_1=1$,
and a term $a_2\eta\, \vbc_{\a}\cdt\vc_{\a}$ with $a_2=1/2$.
We might ask how properties of the theory change if we
treat $a_1$ and $a_2$ as free parameters.
For arbitrary $a_1$ and $a_2$, the effective potential reads
\begin{equation}
  v_{eff} = \frac{\h^2}{2\tl_0}
  + \int^M \frac{d^4p}{(2\p)^4} \left(
          \log[2p^2(p^2+a_1\h)] - 2\log(p^2+a_2\h) \right) \ ,
\label{veffcgh}
\end{equation}
where we have set $\r_0=0$ ({\it c.f.} Eq.~(\ref{veff})).
The $\eta$ gap equation becomes
\begin{equation}
  \frac{\h}{\tl_0} = \int^M \frac{d^4p}{(2\p)^4}
  \left( \frac{2a_2}{p^2+a_2\h} - \frac{a_1}{p^2+a_1\h} \right) \ .
\label{gapeqcgh}
\end{equation}
We see that, except for $a_1/a_2=2$, any other ratio would give rise
to a quadratic divergence, which, in turn, would have to be absorbed
into the bare `t~Hooft coupling $\tl_0= \tg_0^2 N$.  But this cannot happen,
because we know from weak-coupling perturbation theory that the
longitudinal coupling constant $\tg_0$ undergoes only logarithmic
renormalization.  This provides a check on our calculations,
because the solution in the large-$N$ limit has to be consistent with
weak-coupling perturbation theory.

\subsection{\label{erole} Role of eBRST symmetry and mass gap}
The large-$N$ solution of the previous subsection gives rise to what
we may call, in a loose sense, a mass gap.  By this we mean that
the inverse ghost propagator does not vanish at $p^2=0$,
whereas the inverse scalar
propagator vanishes like $p^2$, instead of like $(p^2)^2$.  Explicitly,
we read off from the large-$N$ solution the following propagators
\begin{subequations}
\label{propagate}
\begin{eqnarray}
  \svev{m_i m_j^*}(p) &=& \d_{ij}\,\frac{1}{2p^2(p^2+\eta)}\ ,
\label{propagatea}\\
  \svev{C_{i\a} \bc_{j\b}}(p) &=& \d_{ij}\d_{\a\b}\,\frac{1}{p^2+\eta/2}\ .
\label{propagateb}
\end{eqnarray}
\end{subequations}

In order to appreciate the significance of this result, we recall
that adding the following mass terms to the lagrangian
preserves (on-shell) eBRST invariance \cite{GSMF}
\begin{eqnarray}
  \cl_{\rm mass} &=& 2m_0^2\, \tr\! \left(W_\m^2/(2\tg^2) + \bc C \right)
\label{Smass}\\
  &=& m_0^2 \left( 2 \partial_\m \vm^\dagger \cdt \partial_\m \vm
  + \vbc_{\a}\cdt\vc_{\a}\right) + O(1/\sqrt{N})\ .
\nonumber
\end{eqnarray}
On the second line we used Eq.~(\ref{Vmuphi}) and performed the same rescaling
as in Sec.~\ref{largeN}.
Together with Eq.~(\ref{SN}) this gives rise to the tree-level propagators
\begin{subequations}
\label{treeprop}
\begin{eqnarray}
  \svev{m_i m_j^*}(p) &=& \d_{ij}\,\frac{1}{2p^2(p^2+m_0^2)}\ ,
\label{treepropa}\\
  \svev{C_{i\a} \bc_{j\b}}(p) &=& \d_{ij}\d_{\a\b}\,\frac{1}{p^2+m_0^2}\ .
\label{treepropb}
\end{eqnarray}
\end{subequations}
These propagators respect eBRST symmetry, as can be verified,
for example, by examining the Ward--Takahashi identity
\begin{equation}
  \svev{s'(m_i(x) \bc_j^*(y))} = 0 \ .
\label{WImass}
\end{equation}
Here $s'$ is the modified eBRST transformation constructed in App.~\ref{s'}.%
\footnote{
  We checked the $O(1)$ and $O(\tg^2N)$ terms of this identity in weak-coupling
  perturbation theory.
  Notice that the rescaling $\vm\to \tg\vm$ is to be performed
  in the transformation rules~(\ref{smsc}) and~(\ref{spgh}) as well.
}
We note that, while $m_0=0$ is not protected by eBRST symmetry,
it can nevertheless be shown that no mass terms will be induced to all orders
in weak-coupling perturbation theory, if the tree-level mass vanishes
\cite{GSMF,GZ}.

At face value, comparing the tree-level propagators~(\ref{treeprop})
to the dynamically generated propagators of Eq.~(\ref{propagate}),
reveals a factor two mismatch between the ghost and scalar mass-squared
values.  Furthermore, at the end of the previous subsection
we have seen that this ratio is not incidental, but rather, is
required in order to avoid quadratic divergences.

This state of affairs would seem to imply that eBRST symmetry
is broken spontaneously in the large-$N$ limit.   However,
spontaneous breaking of eBRST would imply the existence of a
massless state with ghost number equal to one,
whereas the solution we found yields a non-vanishing mass for the ghost
field, making it highly unlikely that the spectrum of the theory defined
by the ``truncated'' action~(\ref{SN}) (without the $1/\sqrt{N}$ corrections)
would contain such a Nambu--Goldstone state.   In fact, this action
breaks the modified eBRST symmetry $s'$ ({\it c.f.} App.~\ref{s'}) explicitly, and there is no paradox.   But this
implies that the conclusions we obtained in Sec.~\ref{gapN} cannot
be the full story in the complete theory including $1/\sqrt{N}$ corrections,
because that theory is invariant under $s'$.   This points to limitations
on what we can learn from the large-$N$ framework, to be discussed in
the next subsection.

\subsection{\label{limitation} Limitations of the large-$N$ framework}
Let us revisit the steps that have led to the large-$N$ solution
of Sec.~\ref{gapN}.  First, we write the full lagrangian as
\begin{equation}
  \cl = \cl_N + \cl_1 \ ,
\label{ltot}
\end{equation}
where $\cl_N$ consists of the terms shown explicitly on
the right-hand side of Eq.~(\ref{SN}), while $\cl_1$ corresponds to
the $O(1/\sqrt{N})$ part we have disregarded until now.
As already mentioned at the end of the previous subsection,
a hint that the fate of eBRST symmetry cannot be decided by considering
$\cl_N$ only is that $\cl_N$ and $\cl_1$ are not separately invariant
under the eBRST transformation $s'$, only their sum is.

The existence of the $\cl_1$ part is
unlike what we are used to in the $CP(N)$ model
\cite{DDL1}.  It is possible to reformulate the lagrangian
of the $CP(N)$ model such that it has features similar
to the purely scalar lagrangian of Sec.~\ref{boson}.
With a suitable set of auxiliary and Lagrange-multiplier fields, the action
of the $CP(N)$ model
is bilinear in the unconstrained length-$N$ vector, and this result
is reached without neglecting any terms, regardless of how
they scale formally in large-$N$ counting.

In the $CP(N)$ model it is thus possible to integrate out
the length-$N$ vector exactly.
In the resulting effective action, the running coupling is traded
via dimensional transmutation with a dynamically generated scale $\L$,
and the dependence on any parameters that require renormalization
is eliminated entirely.  We thus have, in principle,
an exact non-perturbative solution of the theory in the large-$N$ limit.
For large, but finite, $N$, a systematic expansion in $1/N$ can be carried out.
Because $N$ is a positive integer,
the expansion parameter $1/N$ does not renormalize.
In other words, the large-$N$ expansion is expected to be a finite
expansion, in contrast with weak-coupling perturbation theory.

Returning to the case at hand,
let us see how far we can get while taking into account the role of $\cl_1$.
We begin by writing the partition function of the
reduced model in the coset-gauge (including the norm constraint) as
\begin{equation}
  Z = \int d\h\, Z(\h) = \int d\h \exp(-W(\h)) \ .
\label{Zeta}
\end{equation}
The (ordinary) integral is over the constant mode of the auxiliary field,
which, as before, we denote by $\h$ for brevity.
The saddle-point equation for $W(\h)$, or the gap equation, is
\begin{equation}
  0 = \frac{\partial W}{\partial\h}
  = \svev{\frac{\partial \cl}{\partial\h}}_\h \ ,
\label{dWeta}
\end{equation}
where $\svev{\cdot}_\h$ denotes the expectation value with respect to
the fixed-$\h$ partition function $Z(\h)$,
and $\cl$ is the full lagrangian of Eq.~(\ref{ltot}).
Explicitly, the gap equation reads
\begin{equation}
  \frac{\h}{\tl_0}
  = \frac{1}{2N} \left( G^{gh}_{i\a i\a}(0) + 4 \tG^\phi_{ii}(0) \right) \ ,
\label{gapexact}
\end{equation}
where the two-point functions are
\begin{subequations}
\label{twopoint}
\begin{eqnarray}
  G^{gh}_{i\a j\b}(x) &=& \svev{C_{i\a}(0) \bc_{j\b}(x)}_\h \ ,
\label{twopointa}\\
  G^\phi_{ij}(x) &=& \svev{m_i(0)\, m_j^*(x)}_\h \ ,
\label{twopointb}\\
  \tG^\phi_{ij}(x) &=&  \bo G^\phi_{ij}(x) \ .
\label{twopointc}
\end{eqnarray}
\end{subequations}
In itself, the gap equation~(\ref{gapexact}) is exact.
The reason is that, by construction,
the auxiliary field $\eta$ occurs only in $\cl_N$,
namely, in the terms explicitly shown on the right-hand side of
Eq.~(\ref{SN}), and not in $\cl_1$.
The question is, however, how much we know about the
two-point functions appearing on the right-hand side.

The momentum-space inverse two-point functions take the form
\begin{subequations}
\label{lowp}
\begin{eqnarray}
  \G^{gh}_{i\a j\b}(p^2) &=& \d_{ij}\d_{\a\b}\,p^2 Z^{gh}(p^2)
  + \S^{gh}_{i\a j\b}(p^2) \ ,
\label{lowpa}\\
  \G^\phi_{ij}(p^2) &=& \d_{ij}\,2(p^2)^2 Z^\phi(p^2) + \S^\phi_{ij}(p^2) \ ,
\label{lowpb}
\end{eqnarray}
\end{subequations}
where $Z^{gh}(p^2)$ and $Z^\phi(p^2)$ are perturbative wave-function
renormalizations.
The additional pieces, $\S^{gh}_{i\a j\b}(p^2)$ and $\S^\phi_{ij}(p^2)$,
contain all the non-perturbative physics, and, by definition,
are the parts that dominate over $p^2$, respectively, $2(p^2)^2$,
in the limit $p^2\to0$.
For large momentum, these pieces must vanish, in order to comply
with the renormalization structure of the theory.  This takes into account
the fact that, as we have already mentioned,
mass terms cannot be induced in perturbation theory \cite{GSMF,GZ}.

The approximations we have applied in Sec.~\ref{gapN} are legitimate
in the regime $p\gg\tL$.
In this regime the running coupling $\tg_r$ is weak.
All corrections to the tree-level propagators, including those
that depend on $\cl_1$, will be down by $\tg_r^2(\m)=\l_r(\m)/N$ with $\m^2=p^2$.
This, in turn, means that they are indeed suppressed by $1/N$.

The situation is qualitatively different for $p\sim\tL$.
Because of the existence of $\cl_1$, unlike in the $CP(N)$ model,
we are bereft of the delicate structure that is needed in order that all of
the non-perturbative effects will be
controlled by, and only by, the unrenormalized small parameter $1/N$.
Close to the scale $\tL$ where it blows up,
the value of the one-loop running `t~Hooft coupling $\tl_r(\m)$ exceeds $N$,
at which point $\tg_r^2(\m)$ becomes of order one.  Thus,
while it is always true that $\tg_r^2(\m)$ is smaller than $\tl_r(\m)$
by $1/N$, the effects of $\tg_r^2(\m)$ are no longer suppressed
when $\tl_r(\m)$ has grown as big as $N$.  This applies, in particular,
to those contributions to the two-point functions that depend on $\cl_1$.
They now reflect the strong dynamics of the longitudinal
coupling at the dynamically generated scale, and all we can say is that
$\S^{gh}_{i\a j\b}(p^2)$ and $\S^\phi_{ij}(p^2)$ scale like $\tL^2$,
respectively, $\tL^4$, for $p\sim\tL$.

In order to make these observations slightly more concrete, and to study their
consequences, we proceed as follows.
First, we will now take Eq.~(\ref{gapsolv}) as the definition of $\tL$ in terms
of the bare parameters of the theory.  Instead of Eq.~(\ref{tL}),
which we no longer expect to be true, our goal will be to derive a more
modest result, namely, that $\sqrt\h$ is still a quantity of order $\tL$.
Since the expectation value $\h$ is a physical observable,
by showing that $\sqrt\h\sim\tL$ we will re-establish that $\tL$
is the dynamically generated scale of the theory.

We begin by splitting the momentum integral implicit
on the right-hand side of Eq.~(\ref{gapexact}) as
\begin{equation}
  \int_0^M = \int_0^{\m} + \int_{\m}^M \ .
\label{intrange}
\end{equation}
The arbitrary scale $\m$ is chosen sufficiently larger than $\tL$
such that, for $p\ge\m$,
the running coupling is weak enough that
the approximations we have made in Sec.~\ref{gapN} are justified.
The contribution of the high-momentum region is equal to $\h/2$ times an
integral whose form is similar
to the right-hand side of Eq.~(\ref{gaplog}), except that the lower limit
is replaced by $\m$, and,
as a result, the neglected terms are indeed suppressed by $1/N$.
As for the low-momentum integral between 0 and $\m$,
all we know at this point is that we should be able to express it
in terms of the available dimensionful quantities of the renormalized theory,
which are the infrared scale $\tL$,
the expectation value $\eta$, and the arbitrary scale $\m$.
The contribution of the low-energy integral can thus be written as $\h$
times $f_{IR}(\sqrt\eta/\tL, \m/\tL)$
with some unknown dimensionless function $f_{IR}(x,y)$.
Isolating the infrared part by moving the contribution
of the high-momentum integral over to the other side,
the gap equation~(\ref{gapexact}) now takes the form
\begin{equation}
  f_{IR}(\sqrt\eta/\tL, \m/\tL)
  = \frac{1}{\tl_0}
    - \frac{1}{2}\int^M_\m \frac{d^4p}{(2\p)^4} \frac{1}{(p^2)^2}
  \equiv \frac{1}{\tl_r(\m)} \ .
\label{solvexact}
\end{equation}
This result is valid up to corrections of order $1/N$,
which are now truly suppressed, because we have been careful to only use
the expansion in $1/N$ in the large-momentum region.

What we have achieved so far, is to rewrite
the exact gap equation such that it no longer depends on
the ultraviolet cutoff $M$ nor on the bare coupling $\tl_0$.
These were traded with the dependence on the running coupling
at the renormalization scale $\m$.  Next, let us study the change
in $f_{IR}$ as we go from $\m$ to some $\m'>\m$.  This change
arises from the contribution of the momentum range $\m\le p \le \m'$.
Up to $1/N$ corrections, once again it is
given by an integral similar to the right-hand side of Eq.~(\ref{gaplog}),
where, this time, the lower limit is $\m$ and the upper limit is $\m'$.
Therefore,
\begin{eqnarray}
  f_{IR}(\sqrt\eta/\tL, \m'/\tL)
  &=& f_{IR}(\sqrt\eta/\tL, \m/\tL)
  + \frac{1}{2}\int^{\m'}_\m \frac{d^4p}{(2\p)^4} \frac{1}{(p^2)^2}
\label{fIRmmp}\\
  &=& \rule{0ex}{3ex}
  f_{IR}(\sqrt\eta/\tL, \m/\tL) + 1/\tl_r(\m') - 1/\tl_r(\m) \ .
\nonumber
\end{eqnarray}
Introducing the subtracted function
\begin{equation}
  f_{IR}^{sub}(\sqrt\eta/\tL, \m/\tL)
  = f_{IR}(\sqrt\eta/\tL, \m/\tL) - 1/\tl_r(\m) \ ,
\label{fIRsub}
\end{equation}
it follows from Eq.~(\ref{fIRmmp}) that $f_{IR}^{sub}$
is in fact independent of the arbitrary scale $\m$,
and thus $f_{IR}^{sub}=f_{IR}^{sub}(\sqrt\eta/\tL)$.
In terms of the subtracted function, the gap equation takes the simple form
\begin{equation}
  f_{IR}^{sub}(\sqrt\eta/\tL) = 0 \ .
\label{gapsub}
\end{equation}
Since this equation involves the single dimensionless variable $\sqrt\eta/\tL$,
it follows that the solution for $\sqrt\h$ is necessarily a quantity
of order $\tL$.%
\footnote{
  In case there is more than one solution, one has to substitute
  each solution back into the effective potential $W(\h)$, to see
  which one is the true minimum.
}
This confirms that $\tL$ is indeed the dynamically generated scale
of the theory.

\subsection{\label{ir} Infrared behavior}
As we have explained in detail in the previous subsection,
the departure of our large-$N$ setup from that of the $CP(N)$ model
turns out to severely limit our ability to solve the theory with its help.
It is only thanks to the logarithmic enhancement of the region
$p\gg\tL$ that we have been able to establish dimensional transmutation.

When we come to the question of how the various symmetries are realized,
evidently the details of $p\sim\tL$ physics matter, because each symmetry
implies a set of Ward--Takahashi identities that are sensitive to
these details.  Lacking sufficient knowledge of the infrared physics,
we are unable to determine which symmetries are spontaneously broken
in the large-$N$ limit, and which ones are not.  With this in mind,
our focus here will be on kinematical considerations,
namely, on how the realization of various symmetries depends
(or, is likely to depend) on the limiting infrared behavior
of various two-point functions.  As already mentioned,
establishing what the infrared behavior really is must await future
studies.

We begin with an observation about the infrared behavior that,
we believe, should be true regardless of the limitations of our large-$N$
setup, and in fact, should be true for any $N\ge2$.
The (fixed-$N$) tree-level propagator of the scalar field $\vm$
of the reduced model is $1/(p^2)^2$.
Just like the familiar massless scalar propagator $1/p^2$
gives rise to infrared singularities in two dimensions, essentially
the same infrared singularities will arise from a $1/(p^2)^2$ propagator
in four dimensions.
A dynamical mechanism that tames these infrared singularities
must therefore be at work.  Within the naive large-$N$ treatment
of Sec.~\ref{gapN}, this role was taken up by the expectation value $\h$.
For $\h>0$, the inverse tree-level propagator of the scalar field
is $2p^2(p^2+\h)$, which vanishes only like $p^2$.
Now, as we have learned in Sec.~\ref{limitation} (see in particular Eq.~(\ref{lowp})),
we actually do not know the two-point functions for $p\sim\tL$.
But, whatever the precise additional contributions from $\cl_1$ may be,
we do not expect them to undo the effect of the non-zero
expectation value for $\h$
and therefore to reinstate a $1/(p^2)^2$ propagator for $p\to0$,
because this would
make the theory infrared singular again.

The remaining question is whether the inverse scalar propagator actually
vanishes like $p^2$, or, rather, tends to a non-zero constant for $p\to0$.
This is closely related to another important question:
does the global $SU(N)$ symmetry break spontaneously to
$SU(N-1)\times U(1)$?
A local order parameter for this symmetry breaking is
\begin{equation}
  \tr \langle  T_0 \tA \rangle_\h
  = -2 \langle z^\dagger T_0 z \rangle_\h  \ ,
\label{vevA}
\end{equation}
where $\tA$ is defined in Eq.~(\ref{wAtilde}),
and $T_0 = \tT_0 - (N-2)/N$ is the traceless part of $\tT_0$.
Because the coset generators do not commute with $\tT_0$,
a massless pole in the $\vm$ propagator would allow
the identification of the degrees of freedom of this field
with the $2(N-1)$ NGBs.
The alternative, an inverse propagator for $\vm$ that does not vanish
for $p\to0$, would seem to rule out the spontaneous breaking of $SU(N)$.
First, this would clearly exclude the possibility that the $\vm$ scalar field
itself is a NGB.  Moreover, if the scalar propagator falls
off exponentially, it is hard to conceive of a scalar bound state
that would exhibit a power law fall-off, as would be required for
a NGB.

We next turn to the eBRST symmetry.  An example of a Ward--Takahashi
identity for this symmetry was given in Eq.~(\ref{WImass}),%
\footnote{
  The explicit form is not very illuminating because of the complicated form
  of the $s'$ transformation, see App.~\ref{s'}.
}
which implies a relation between $G^\phi_{ij}$ and $G^{gh}_{i\a j\b}$.
Whether or not eBRST is broken
spontaneously is closely related to the question of whether or not
the ghosts pick up a non-zero mass.  If eBRST symmetry does break
spontaneously, this would necessitate the existence of a massless state
with ghost number equal to one.
In the original formulation of the reduced model
before fixing the coset gauge, the original eBRST transformation $s$
commutes with both the local $H=SU(N-1)\times U(1)$ and the global $G=SU(N)$
transformations.  Because the ghost field $C$ transforms non-trivially under the
local $H$ symmetry, it cannot couple to the eBRST
current as a fundamental field.  However, the eBRST current could couple
to a composite operators with ghost number equal to one.
Since any such composite operator has at least one ghost-field constituent,
it is highly unlikely that any bound state
with a ghost-field constituent could admit a power-law fall-off
if the ghost-field propagator itself falls off exponentially
(in the coset gauge).

We conclude that a massive ghost field clearly favors
unbroken eBRST symmetry.  We point out that,
in close analogy with the scalar field, the naive large-$N$ treatment
of Sec.~\ref{gapN} does give rise to a non-zero ghost mass.
It is difficult to see how
the unaccounted for effects of $\cl_1$ would conspire to precisely
cancel this dynamically induced mass.

Finally, similar considerations apply to the $SL(2,R)$ symmetry (that has
the ghost-number charge among its generators, see Sec.~\ref{basic}).
This symmetry acts only on the ghosts, and therefore, an order parameter
for its breaking would have to be a ghost bilinear that is not a
singlet under $SL(2,R)$.  This includes
as a special case the auxiliary field $\r_0$, which is a ghosts bilinear
by its equation of motion.
Once again, the dynamical generation of a non-zero ghost mass
would seem to rule out the spontaneous breaking of $SL(2,R)$.

\section{\label{conclusion} Conclusion}
In this paper we studied the reduced model of $SU(N)/(SU(N-1)\times U(1))$
equivariantly gauge-fixed theory, in the large-$N$ expansion.
Our main motivation for this study comes from the speculative scenario
put forward in Ref.~\cite{GSMF}, which we reviewed
in some detail in the introduction.
The reduced model describes the longitudinal sector of the
equivariantly gauge-fixed theory, and, like the transverse sector,
it is controlled by an asymptotically free coupling \cite{GSb}.
According to our scenario, depending on the initial values of the
transverse and longitudinal couplings at the cutoff,
the longitudinal sector can become strongly coupled at an infrared scale $\tL$
where the transverse coupling is still weak.
The resulting dynamics would be that of a novel Higgs--Coulomb phase (see Fig.~\ref{phasediag}).

Because the transverse sector is weak, and plays a spectator
role in the dynamics of the novel phase, the
key questions can be studied within the reduced model,
which, we recall, corresponds to the $g_0=0$ boundary of the phase diagram.
The most exciting scenario is that the novel phase is connected
to the gaussian fixed point where the continuum limit is to be taken,
as shown in the right panel of Fig.~\ref{phasediag}.
The conjecture is that
dimensional transmutation is accompanied by the spontaneous breaking of
the global $SU(N)$ symmetry of the reduced model down to $SU(N-1)\times U(1)$,
while none of the other symmetries are broken.  The symmetry breaking
$SU(N)\to SU(N-1)\times U(1)$ would give rise to $2(N-1)$ NGBs,
that will then generate a mass of order $g\tL$ for
the coset gauge fields of the full theory,
where $g\ll 1$ is the usual (transverse)
running coupling at the scale $\tL$.  The photon of the unbroken $U(1)$
would stay massless, whereas the unbroken $SU(N-1)$ theory would
eventually become strong, and confine at a much lower scale
$\tL\exp[-c/((N-1)g^2)]$, where $c = 48\pi^2/11$.
Finally, if both eBRST and $SL(2,R)$ (which include the ghost-number
symmetry) are not broken spontaneously, there is hope that the new phase
might be unitary \cite{KO}.

At a more technical level, the
motivation for using large-$N$ methods comes from the
observation that the number of coset degrees of freedom
grows linearly with $N$.  Theories exhibiting a similar linear growth are often
solvable in closed form in the large-$N$ limit, with a particularly relevant
example being the two-dimensional $CP(N)$ model \cite{DDL1}.

Using the large-$N$ reformulation of the reduced model developed
in Sec.~\ref{largeN}, we have found an approximate solution of the large-$N$
gap equation in Sec.~\ref{gapN}, and we showed that dimensional transmutation
takes place.  Unfortunately, as explained in detail in Sec.~\ref{limitation},
the large-$N$ framework turns out to be significantly more
involved than in the $CP(N)$ model.  In a nutshell, the dynamics is
controlled not only by the unrenormalized small parameter $1/N$,
but also, separately, by the running coupling.
While it is true, by definition, that the ratio of the ordinary coupling
and the `t~Hooft coupling is $\tg_r^2/\tl_r=1/N$,
the ordinary running coupling $\tg_r^2$ is no longer
a parametrically small quantity when the running `t~Hooft coupling
itself becomes as large as $N$.  Under these circumstances,
even establishing
dimensional transmutation requires careful consideration.

As might be expected, the dynamically generated scale $\tL$
is the same as the one occurring in the solution of the
one-loop renormalization-group equation.  This result
rules out an infrared attractive fixed point in the large-$N$ limit;
the low-energy physics of the reduced model is not conformal,
at least for large $N$.

Because of the complications of the large-$N$ framework, we have not been able
to determine the infrared spectrum, and thus we were unable to answer
the question whether dynamical symmetry breaking occurs,
and if so,  what its pattern would be.  For a discussion of the relevant questions,
we refer to Sec.~\ref{ir}.  If we could show that the
strong dynamics of the reduced model drives the spontaneous breaking of
$SU(N)$ to $SU(N-1)\times U(1)$, this would confirm
the scenario shown in the right panel of Fig.~\ref{phasediag}.
With the current state of affairs, our knowledge of the phase diagram remains
rather limited.  We have not ruled out a scenario where none of the symmetries
of the reduced model are broken spontaneously by its strong dynamics;
and thus we did not prove the speculative scenario in which
the strong longitudinal dynamics affects the physics in the transverse sector.

Still, it is encouraging that two rather different lines of investigations,
the mean-field study of Ref.~\cite{GSMF}, and, now, large $N$,
both provide evidence that is compatible with
the exciting scenario shown in the right panel of Fig.~\ref{phasediag}.

In the future, we plan to return to the study of equivariantly gauge-fixed
theories using other non-perturbative methods, including,
in particular, numerical techniques.  We note that, should
the existence of the novel phase with $SU(N)\to SU(N-1)\times U(1)$
symmetry breaking be established via Monte-Carlo simulations,
our large-$N$ analysis in this paper would provide strong evidence
that it is connected to the continuum limit, and not a lattice artifact,
at least for large $N$.

\vspace{3ex}
\noindent {\bf Acknowledgments}
\vspace{3ex}

MG thanks the School of Physics and Astronomy of Tel Aviv University,
the Department of Physics at the Universit\`a
``Tor Vergata'' and IFAE at the Universitat Aut\`onoma de Barcelona,
and YS thanks the Department of Physics and Astronomy of San Francisco
State University for hospitality.
MG is supported in part by the US Department of Energy, and
YS is supported by the Israel Science Foundation under grants no.~423/09 and~449/13.

\appendix
\section{\label{s'} Modified eBRST transformation}
When the scalar field $\phi$ is parametrized as in Eq.~(\ref{phiHGH}),
the eBRST transformation rule~(\ref{sphi}) mixes $\phi_H$ and $\phi_{G/H}$.
In order to have a symmetry that preserves the coset gauge $\phi=\phi_{G/H}$,
the eBRST transformation rules of all fields have to be modified
by a compensating $H$ transformation.
This can be worked out as follows.

We begin by considering the $H$-invariant operator
\begin{equation}
  \tA = \phi^\dagger\tT_0\phi \ .
\label{tildeA}
\end{equation}
This operator is the natural generalization
of an operator with the same name that we introduced in Ref.~\cite{GSMF},
and it can serve as an order parameter for $SU(N)\to SU(N-1)\times U(1)$
symmetry breaking (see Sec.~\ref{ir}).
First using the representation~(\ref{whi}) for $\phi$
and the unitarity relations~(\ref{unirels}) one has
\begin{subequations}
\label{wAtilde}
\begin{equation}
  \tA = I_N - 2\cp \ ,
\label{wAtildea}
\end{equation}
where $\cp$ is defined in Eq.~(\ref{Pz}).
Moreover, because $\tA$ is $H$-invariant, it is independent of $\phi_H$,
hence $\tA(\phi)=\tA(\phi_{G/H})$.  In other words, $\tA$ can be expressed
in terms of the degrees of freedom in $\phi_{G/H}$ only,
\begin{equation}
  \tA = I_N - 2
      \left(\begin{array}{c}
        \vm \\
        c
      \end{array}\right)
      \left(\begin{array}{cc}
        \vm^\dagger & c
      \end{array}\right) \ ,
\label{wAtildeb}
\end{equation}
\end{subequations}
where now $c$ is real.
In addition, when calculating the eBRST transformation of $\tA$,
it must be possible to consistently interpret the result as arising from
a new transformation, $s'$, that acts only on the degrees of freedom
in $\phi_{G/H}$.  Explicitly,
\begin{eqnarray}
  s \tA &=& i \phi^\dagger [C,\tT_0] \phi
\label{ebrstz1}\\
  &=& \rule{0ex}{6ex}
  i\left( \begin{array}{cc}
    \vm\vc^\dagger(P_\perp+cP_\parallel)-(P_\perp+cP_\parallel)\vc\vm^\dagger &
    -c(P_\perp+cP_\parallel)\vc-\vm(\vc^\dagger\cdt\vm) \\
    c\vc^\dagger(P_\perp+cP_\parallel)+(\vm^\dagger\cdt\vc)\vm^\dagger &
    c(\vm^\dagger\cdt\vc-\vc^\dagger\cdt\vm)
  \end{array}\right)
\NON
  &\equiv& \rule{0ex}{5ex} -2 s'
      \left(\begin{array}{c}
        \vm \\
        c
      \end{array}\right)
      \left(\begin{array}{cc}
        \vm^\dagger & c
      \end{array}\right) \ .
\nonumber
\end{eqnarray}
On the second line, we have evaluated the eBRST variation on $\phi=\phi_{G/H}$.
The projectors are
\begin{eqnarray}
\label{wproj}
\cp_\perp &=& I_{N-1} - \hv\hv^\dagger\ ,\\
\cp_\parallel &=& \hv\hv^\dagger\ .
\label{wprojll}
\end{eqnarray}
We may now read off the transformation rules
\begin{subequations}
\label{smsc}
\begin{eqnarray}
  s' c  &=& \frac{i}{4}\left(\vc^\dagger\cdt\vm-\vm^\dagger\cdt\vc\right)\ ,
\label{smsca}\\
  s'\vm &=& \frac{i}{2}(P_\perp+cP_\parallel)\vc
  +\frac{i}{4c}\,\vm\left(\vm^\dagger\cdt\vc+\vc^\dagger\cdt\vm\right)\ .
\label{smscb}
\end{eqnarray}
\end{subequations}
We first obtained the $s'c$ rule by comparing the lower-right block entry
of the matrices on the second and third lines of Eq.~(\ref{ebrstz1}).
Knowing $s'c$, we inferred $s'\vm$ from the off-diagonal blocks.
Finally, the upper-left block provides a consistency check.
Equation~(\ref{smsc}) implies that $s'(z^\dagger z) = 0$,
consistent with the constraint~(\ref{znorm}).
Also, the right-hand side of Eq.~(\ref{smsca}) is (formally) real,
respecting the reality of $c$ according to Eq.~(\ref{wcomp}).

With Eq.~(\ref{smsc}) in hand,
we calculate the action of $s'$ on $\phi_{G/H}$ using Eq.~(\ref{phiGHmc}) again,
finding
\begin{subequations}
\label{trans}
\begin{equation}
  s' \phi_{G/H} = \frac{i}{2} \left( \begin{array}{cc}
  \ca & \cb \\
  \cc & \cd
  \end{array}\right)  \ ,
\label{transa}
\end{equation}
where
\begin{eqnarray}
  \ca &=& -\frac{1}{1+c}\left(\vc\vm^\dagger-\vm\vc^\dagger
        +\half(c-1)P_\parallel(\vm^\dagger\cdt\vc-\vc^\dagger\cdt\vm)\right)\ ,
\label{transb}\\
  \cb &=& -(P_\perp+c P_\parallel)\vc
        -\frac{1}{2c}\vm(\vm^\dagger\cdt\vc+\vc^\dagger\cdt\vm)\ ,
\label{transc}\\
  \cc &=& -\vc^\dagger(P_\perp+c P_\parallel)-\frac{1}{2c}
        (\vm^\dagger\cdt\vc+\vc^\dagger\cdt\vm)\vm^\dagger \ ,
\label{transd}\\
  \cd &=& \half(\vc^\dagger\cdt\vm-\vm^\dagger\cdt\vc) \ .
\label{transe}
\end{eqnarray}
\end{subequations}
By contrast, using the normal eBRST transformation~(\ref{sphi}), we find
\begin{equation}
\label{normals}
s\phi_{G/H}=-i C\phi_{G/H} =-\frac{i}{2}\left( \begin{array}{cc}
\vc\vm^\dagger & c\vc \\
\vc^\dagger\left(1-\frac{1}{1+c}\vm\vm^\dagger\right) & -\vc^\dagger\cdt\vm
 \end{array}\right) \ .
\end{equation}
It is no surprise that Eq.~(\ref{normals}) does not agree with Eq.~(\ref{trans}).
Indeed we have
\begin{equation}
  s \phi = s (\phi_H \phi_{G/H})
  = (s \phi_H)\phi_{G/H} + \phi_H(s \phi_{G/H})\ ,
\label{stotphi}
\end{equation}
and so the discrepancy arises from the missing term $(s \phi_H)\phi_{G/H}$.
In the limit where $\phi_H$ is close to the identity matrix,
the missing term takes the form of an infinitesimal $H$ transformation
with a Grassmann parameter that acts on $\phi_{G/H}$.
The conclusion is that we must have
\begin{equation}
\label{spsrel}
s'\f_{G/H}=s\f_{G/H}+\d_H \, \f_{G/H}\ .
\end{equation}
Here $\d_H$ is a compensating $H$ transformation that reinstates
the coset gauge, which can be parametrized as
\begin{equation}
\label{Htrans}
\d_H \, \f_{G/H} =
i\left( \begin{array}{cc}
\ca\left(1-\frac{1}{1+c}\vm\vm^\dagger\right)& -\ca\vm\\
\th\vm^\dagger &
\th c
 \end{array}\right) \ ,
\end{equation}
where the $(N-1)\times(N-1)$ matrix $\ca$ is (formally) hermitian.
We solve for $\ca$ and $\th$ by requiring Eq.~(\ref{spsrel}) to hold, finding
\begin{subequations}
\label{Hsol}
\begin{eqnarray}
\ca&=&\half\frac{1}{1+c}\left(\vc\vm^\dagger+\vm\vc^\dagger+
\frac{1-c}{2c}P_\parallel\left(\vm^\dagger\cdt\vc+\vc^\dagger\cdt\vm\right)\right)\ ,
\label{Hsola}\\
\th&=&-\frac{1}{4c}\left(\vm^\dagger\cdt\vc+\vc^\dagger\cdt\vm\right)\ .
\label{Hsolb}
\end{eqnarray}
\end{subequations}
Once again, we have an over-constrained system, which provides new
consistency checks.%
\footnote{
  For example, the two lower blocks of Eq.~(\ref{Htrans})
  both give rise to the solution~(\ref{Hsolb}) for $\th$.
}
Observe that $\tr(\ca)+\th=0$, consistent with the tracelessness of
all generators of $H=SU(N-1)\times U(1)$.

It is now straightforward to obtain the $s'$ transformation rules
of the ghost fields as well, which read
\begin{subequations}
\label{spgh}
\begin{eqnarray}
s'\vc&=&i(\ca-\th)\vc\ ,
\label{spgha}\\
s'\vc^\dagger&=&i\vc^\dagger(\ca-\th)\ ,
\label{spghb}\\
s'\vbc&=&-\frac{2i}{\tg^2}\left((P_\perp+c P_\parallel)D^2\vm-(D^2 c)\vm \right)
+i(\ca-\th)\vbc\ ,
\label{spghc}\\
s'\vbc^\dagger&=&\frac{2i}{\tg^2}\left((D^2\vm)^\dagger(P_\perp+c P_\parallel)
-(D^2 c)^*\vm^\dagger\right)+i\vbc^\dagger(\ca-\th) \ .
\label{spghd}
\end{eqnarray}
\end{subequations}
The abelian covariant derivative is defined in Eq.~(\ref{Da}).
Note that $sC=0$, and therefore Eqs.~(\ref{spgha}) and~(\ref{spghb}) arise
from the $\d_H$ part only.  On the right-hand side of Eq.~(\ref{spghb}),
the factor of $i$ (rather than $-i$) originates from the fact
that $\ca$ and $\th$ are anti-commuting (a similar comment applies to
the relevant term in Eq.~(\ref{spghd})).

\section{\label{MAG} Comparison with maximal abelian gauge}
In the special case $N=2$, the partial gauge fixing of $SU(2)$ down to $U(1)$
is a common feature of our eBRST framework and that
of the familiar maximal abelian gauge
(MAG) \cite{mag,Sorella,Kondo,Schaden,Mendes,Suganuma,Bornyakov}.
While $N=2$ is the furthest one can be away from large $N$,
the common coset structure make it interesting
to compare the two approaches.

At the algebraic level, the main difference between the eBRST framework and MAG
is that, in MAG, the coset gauge fixing gives rise to
a standard BRST symmetry, while in our case it is an equivariant
BRST symmetry.
Let us first consider perturbation theory.  For eBRST, this
was discussed in Refs.~\cite{nachgt,MS1,GSMF,GSb,FF}.  In this context there are
many similarities. First, in a perturbative setup,
one must always accompany the gauge fixing of the coset $SU(2)/U(1)$
by a further gauge fixing of the remaining $U(1)$.  The two frameworks
also share an $SL(2,R)$ symmetry that acts on the ghost sector,
as well as the same one-loop beta function for the longitudinal coupling
$\tg$ \cite{GSb,Sorella}.

Non-perturbatively, the key difference is that the MAG gauge-fixing
action cannot be put on the lattice, because, by Neuberger's theorem
\cite{HNnogo}, its BRST symmetry will give rise to a partition function
that vanishes identically.  We note that
this was the main motivation for
introducing eBRST gauge fixing in the first place \cite{nachgt,MS1}.\footnote{%
  We stress that, in the eBRST framework, the ``auxiliary'' gauge fixing
  of the remaining $U(1)$ is not done on the lattice.  It is only needed
  in order to make contact with weak-coupling lattice perturbation theory, in the same
  manner that gauge fixing is needed for perturbation theory
  in the usual gauge-invariant lattice formulation.
}

On the lattice, MAG usually has the following operational
meaning.  First, an ensemble of configurations is generated using
the usual gauge-invariant Boltzmann weight.  Then each configuration
is rotated by a gauge transformation such that the functional
\begin{equation}
  -\sum_{x,\m} \tr( U_{x,\m} \tT_0 U_{x,\m}^\dagger \tT_0 ) \ ,
\label{lattmag}
\end{equation}
reaches a (local) minimum.\footnote{%
  Because the functional~(\ref{lattmag}) is invariant under $U(1)$
  gauge transformations, the same is true for the manifold of local minima.
  We note that the same functional occurs in Eq.~(A.3) of Ref.~\cite{GSMF}.
}
This state of affairs means that the connection between
analytic \cite{mag,Sorella,Kondo,Schaden} and
numerical \cite{Mendes,Suganuma,Bornyakov} studies of MAG
is somewhat subtle.

In the eBRST framework,
the lattice action is simply a suitable discretization
of the continuum action.\footnote{%
  Lattice discretizations have been worked out in detail
  in Refs.~\cite{nachgt,GSMF}.
}
Ghost fields exist on the lattice,
and the ghost propagator is defined as the two-point function of the
ghost fields.
By contrast, in the lattice MAG approach there are no ghost fields.
In this case one usually identifies the ghost propagator
with the inverse of the hessian matrix
obtained by differentiating the functional~(\ref{lattmag}) twice with respect
to a gauge variation.
The eBRST framework and the MAG approach thus give rise to
qualitatively different ghost propagators.  Because of the minimization
of the functional~(\ref{lattmag}) involved in its implementation,
in the MAG approach the ghost propagator is the inverse
of a non-negative matrix.  By contrast, in the eBRST framework, the ghost
two-point function can have both positive and negative eigenvalues.
While the full impact of these qualitative differences is not known,
it is plausible that this could lead to dynamical differences, in particular
for the phase diagram.

We next turn to a comparison of the dynamical predictions
obtained in the two cases.
Originally, MAG was introduced to study abelian dominance,
and the monopole-condensation picture of confinement \cite{mag}.
More recent studies mostly focus on the dynamical generation of a mass gap
for the gluons and/or for the ghosts.
A word of caution is that the Dyson--Schwinger gap equation,
which is common in analytic studies of MAG \cite{Sorella,Kondo},
is not a controlled approximation.\footnote{%
  We are not aware of any large-$N$ study of MAG.
}

With these reservations in mind, in MAG one finds that a mass gap
is generated, and that the scale characterizing the mass gap is $\L$,
\ie, the scale obtained via dimensional transmutation from the usual
gauge coupling.  This is accompanied by the formation of various
two-body condensates, including one that spontaneously breaks the
$SL(2,R)$ symmetry \cite{Sorella}.
These results appear to be, at least qualitatively,
in agreement with numerical lattice studies \cite{Mendes,Suganuma,Bornyakov}.

In our own work---both in Ref.~\cite{GSMF} and in this paper---we have focused
on the part of the phase diagram where the running longitudinal coupling
$\tg$ becomes strong while the usual gauge coupling $g$ is still weak.
Correspondingly, we conjecture,
based on the findings of Ref.~\cite{GSMF} and the
present paper, that the non-perturbative dynamics
is not characterized by the usual confinement scale $\L$,
but rather, by the new scale $\tL$
obtained from the longitudinal coupling $\tg$ via dimensional transmutation.
Now, in our conjectured phase diagram, Fig.~\ref{phasediag},
the confinement phase (phase A) is characterized by the scale $\L$,
whereas the Higgs--Coulomb phase (phase B) is characterized by the scale $\tL$,
which, in phase B, satisfies $\tL\gg\L$.
Regardless of the accuracy of the various predictions of both
frameworks might be,
it thus appears that the MAG approach is geared toward phase A,
the confinement phase, while our interest is focused on the
possible existence of phase B, a Coulomb--Higgs like phase.

\vspace{5ex}

\end{document}